\documentclass{JHEP3}
\usepackage{amsmath,amssymb,epsfig,cite}

\title{Gravitino Dark Matter Scenarios with Massive Metastable Charged Sparticles at the LHC}

\author{John~R.~Ellis\\ 
Theory Division, Physics Department, CERN, CH-1211 Gen\`eve, Switzerland \\
E-mail: \email{John.Ellis@cern.ch}}

\author{Are~R.~Raklev\\ 
Department of Physics and Technology, University of Bergen, N-5007
Bergen, Norway\\
and Theory Division, Physics Department, CERN,
CH-1211 Gen\`eve, Switzerland \\ E-mail: \email{Are.Raklev@cern.ch}}

\author{Ola~K.~{\O}ye\\ 
Department of Physics and Technology, University of Bergen, N-5007
Bergen, Norway\\ E-mail: \email{oye@ift.uib.no}}

\abstract{We investigate the measurement of supersymmetric particle
masses at the LHC in gravitino dark matter (GDM) scenarios where the
next-to-lightest supersymmetric partner (NLSP) is the lighter scalar
tau, or stau, and is stable on the scale of a detector. Such a massive
metastable charged sparticle would have distinctive Time-of-Flight
(ToF) and energy-loss ($dE/dx$) signatures. We summarise the
documented accuracies expected to be achievable with the ATLAS
detector in measurements of the stau mass and its momentum at the
LHC. We then use a fast simulation of an LHC detector to demonstrate
techniques for reconstructing the cascade decays of supersymmetric
particles in GDM scenarios, using a parameterisation of the detector
response to staus, taus and jets based on full simulation
results. Supersymmetric pair-production events are selected with high
redundancy and efficiency, and many valuable measurements can be made
starting from stau tracks in the detector. We recalibrate the momenta
of taus using transverse-momentum balance, and use kinematic cuts to
select combinations of staus, taus, jets and leptons that exhibit
peaks in invariant masses that correspond to various heavier sparticle
species, with errors often comparable with the jet energy scale
uncertainty.}
 \preprint{CERN-PH-TH/2006-135}

\begin{document}

%%%%%%%%%%%%%%%%%%%%%%%%%%%%%%%%%%%%%%%%%%%%%%%%%%%%%%%%%%%%%%%%%%%%%%
\section{Introduction}\label{sect:intro}
%%%%%%%%%%%%%%%%%%%%%%%%%%%%%%%%%%%%%%%%%%%%%%%%%%%%%%%%%%%%%%%%%%%%%%
One of the main motivations for supersymmetry is the existence of a
natural candidate for cold dark matter
(CDM)~\cite{Goldberg:1983nd,Ellis:1983ew} in models in which R-parity
is conserved, thus making the lightest supersymmetric partner (LSP)
stable. Most detector studies of supersymmetric models have focused on
the constrained minimal supersymmetric standard model (CMSSM), in
which the soft supersymmetry-breaking parameters: the gaugino masses
$m_{1/2}$, the scalar masses $m_0$ and the trilinear parameters $A_0$,
are each assumed to be universal at some high scale, and the LSP is
assumed to be the lightest neutralino. With the precision measurement
of the CDM density obtained by combining data from the WMAP experiment
with other cosmological data
\cite{Bennett:2003bz,Spergel:2003cb}, and taking into account accelerator
constraints, the CMSSM parameter space has become quite restricted
\cite{Battaglia:2003ab}. Collider signatures of these models
typically involve missing energy from escaping neutralinos.

However, there is another plausible CDM candidate in R-conserving
supersymmetric models, namely the supersymmetric partner of the
graviton, the
gravitino~\cite{Ellis:1984eq,Moroi:1993mb,Ellis:1995mr,Bolz:1998ek,
Gherghetta:1998tq,Asaka:2000zh,Bolz:2000fu,Fujii:2002yx,Fujii:2002fv,
Buchmuller:2003is,gdm,Feng:2003xh,Feng:2004zu,Feng:2004mt}. The
gravitino mass, $m_{3/2}$, is poorly constrained by accelerator
experiments and the astrophysical and cosmological constraints are
very different from those on the neutralino. In gravitino dark matter
(GDM) scenarios the next-to-lightest supersymmetric partner (NLSP) has
a very long lifetime, particularly in gravity-mediated models of
supersymmetry breaking, in which the gravitino's couplings to other
particles are suppressed by an inverse power of the Planck mass. In
the case of a neutralino NSLP and an arbitrary gravitino mass, the
allowed parameter space of CMSSM is expanded and the classic
supersymmetric signature of missing energy at colliders remains. Here
we study instead the intriguing possibility of a charged NLSP, in
particular the case where the NSLP is the lighter stau,
$\tilde\tau_1$. In gravity-mediated models with R conservation, the
stau would be stable on the scale of a detector, and every
supersymmetric event would contain a pair of massive metastable
charged leptons. Because of their large mass and low velocities, these
would have distinctive Time-of-Flight (ToF) and energy-loss ($dE/dx$)
signatures. Moreover, most supersymmetric events in such stau NLSP
scenarios also contain a pair of $\tau$ leptons as well as energetic
jets and possibly other leptons~\cite{DeRoeck:2005bw}.

In Section~\ref{sect:benchmarks} we review the properties of three
benchmark points for GDM models with a stau NLSP, which were first
proposed in~\cite{DeRoeck:2005bw}. Subsequently, in
Section~\ref{sect:MC}, we summarise the documented expectations for
the ATLAS detector response to the metastable $\tilde\tau_1$ with
limited integrated luminosity, focusing on the accuracy achievable at
the LHC in measurements of the stau mass and momentum for these
benchmark points, taking into account trigger information and looking
at cuts to reject possible backgrounds. In Section~\ref{sect:fastsim}
we use a parametrised fast simulation to investigate the capabilities
of ATLAS to measure the masses of other supersymmetric particles in
GDM models by reconstructing supersymmetric cascade decays. For this
we use as building blocks the staus themselves, the accompanying
$\tau$ leptons, whose full momenta we can reconstruct using transverse
momentum information, hadronic jets and any accompanying charged
leptons. In the benchmark scenario $\epsilon$, with a relatively light
spectrum, many different sparticle species may be reconstructed in
this way and their masses measured with a high accuracy. In the
heavier benchmark scenarios $\zeta, \eta$, fewer sparticles can be
reconstructed and with worse accuracy, even at the fairly high
integrated luminosity considered. Finally, in
Section~\ref{sect:conclusion} we draw our main conclusions.

%%%%%%%%%%%%%%%%%%%%%%%%%%%%%%%%%%%%%%%%%%%%%%%%%%%%%%%%%%%%%%%%%%%%%%
\section{The GDM Benchmark Points}\label{sect:benchmarks}
%%%%%%%%%%%%%%%%%%%%%%%%%%%%%%%%%%%%%%%%%%%%%%%%%%%%%%%%%%%%%%%%%%%%%%
Three GDM benchmark points with a $\tilde\tau_1$ NLSP, named
$(\epsilon,\zeta,\eta)$, were proposed in~\cite{DeRoeck:2005bw}. These
were formulated in the framework of minimal supergravity (mSUGRA)
models~\cite{Nilles:1983ge}, in which the gravitino mass is fixed
equal to the universal soft supersymmetry-breaking masses of scalar
particles at a GUT input scale: $m_{3/2}=m_0$, and there is a simple
relation between the soft trilinear and bilinear parameters:
$A_0=B_0+m_0$. This relationship can be used to fix the ratio of the
Higgs vacuum expectation values $\tan\beta$ from the electroweak
vacuum conditions. Further, the value $A_0=(3-\sqrt{3})\,m_0$ found in
the Polonyi model of supersymmetry breaking in a hidden sector is
assumed~\cite{Polonyi:1977pj}. The resulting allowed GDM parameter
space may be displayed as a region in the $(m_{1/2},m_0)$ plane,
subject to theoretical, phenomenological and cosmological constraints.

This region is restricted, in particular, by the requirement of
maintaining the cosmological concordance between the values of the
baryon-to-entropy ratio inferred from the cosmic microwave background
radiation and from astrophysical light-element abundances, which
constrain NLSP decays and cut the parameter space down to a wedge in
the $(m_{1/2},m_0)$ plane, throughout which the $\tilde\tau_1$ is the
NLSP and is metastable. The shape of this wedge, taken
from~\cite{DeRoeck:2005bw}, is shown in Fig.~\ref{fig:scan1} as the
area restricted to lie below the upper black line. The $\epsilon$
benchmark point is a collider-friendly point at low values of
$m_{1/2}$ and $m_0$ sitting near the apex of the
cosmologically-allowed wedge.  It features relatively large production
cross sections for supersymmetric events at the LHC, as seen in
Table~\ref{table:xsec}. The $\zeta$ and $\eta$ points represent more
challenging scenarios at higher values of $m_{1/2}$, and hence with
lower cross sections. The point $\zeta$ lies at the boundary of the
cosmologically-allowed area, where the NLSP lifetime exceeds $10^6$~s,
whereas $\eta$ is characteristic of models with a relatively low
$\tilde\tau_1$ lifetime $\sim 10^4$~s, shown as the lower black line
in Fig.~\ref{fig:scan1}, below which additional constraints due to
hadronic interactions in the early universe become
important~\cite{Reno:1987qw,Dimopoulos:1988ue,Kohri:2001jx}. A recent
analysis indicates that earlier constraints were overly restrictive,
so that the point $\eta$ lies well within the allowed
region~\cite{Steffen:2006hw}.  None of these three benchmark points
give the full amount of CDM required by WMAP, if one takes into
account only the gravitino abundance due to decays of the NLSP after
freeze-out. The missing CDM could be provided by additional gravitino
production mechanisms in the early Universe, or other components of
CDM such as axions.

\begin{TABLE}{
\begin{tabular}{|c||cccccc|} \hline
Model & $\sigma(\tilde\tau_1\tilde\tau_1^*)$ & $\sigma(\tilde g \tilde g)$ & $\sigma(\tilde g\tilde q)$ & $\sigma(\tilde q\tilde q)$ & $\sigma(\tilde q\tilde q^*)$ & $\sigma(\tilde\chi^0_2 \tilde\chi^\pm_1)$ \\ \hline
$\epsilon$ & $0.0242$ & $0.220$ & $1.36$ & $0.755$ & $0.445$ & $0.114$\\
$\zeta$ & $0.00124$ & $0.000194$ & $0.00391$ & $0.00851$ & $0.00148$ & $0.00229$ \\
$\eta$ & $0.00157$ & $0.000195$ & $0.00394$ & $0.00859$ & $0.00150$ & $0.00229$ \\ \hline
\end{tabular}
\caption{The main supersymmetric production cross sections at the LHC, in pb,
including Drell-Yan production of $\tilde\tau_1$ pairs, pair
production of strongly-interacting sparticles and associated
production of neutralinos and charginos, computed to NLO for the GDM
benchmarks using {\tt Prospino2}~\cite{Beenakker:1996ed}. Note that
these differ from those given in~\cite{DeRoeck:2005bw}, mainly because of
the larger gluino and squark masses obtained from {\tt ISAJET} (see
below) and partly because we use different parton density
functions (see Section~\ref{sect:MC}).}
\label{table:xsec}}
\end{TABLE}

For the purposes of this paper, the effective masses of sparticles at
the electroweak scale for these benchmark points were calculated by
running the universal high-scale masses down to low scales using {\tt
ISAJET~7.69}~\cite{Baer:1999sp}. This gives somewhat larger masses
than found in~\cite{DeRoeck:2005bw} using the {\tt SSARD} code, in
particular for the strongly-interacting sparticles. Both the GUT-scale
input parameters and the resulting physical masses are given in
Table~\ref{table:benchmarks} (as in \cite{DeRoeck:2005bw}, we assume
sign$(\mu)=+$ and $m_t=178$~GeV for all three points).  The decay
widths and branching ratios of the supersymmetric particles were
re-calculated from these masses using the decay code {\tt
SDECAY~1.1a}~\cite{Muhlleitner:2003vg}, except for the three-body
decays of the first- and second-generation right-handed sleptons,
which are not included in {\tt SDECAY}, and whose decays instead are
taken from {\tt ISAJET}.

%\vspace{\stretch{1}}

\begin{TABLE}{
\begin{tabular}{|c||r|r|r|}
\hline
Model & $\epsilon$ & $\zeta$ & $\eta$ \\ \hline
$m_{1/2}$ & 440 & 1000 & 1000 \\
$m_0$ & 20 & 100 & 20 \\
$\tan{\beta}$ & 15 & 21.5 & 23.7 \\
%sign($\mu$) & $+$ & $+$ & $+$ \\
$A_0$ &-25 &-127 &-25 \\ \hline
%$m_t$ 	       & 178 & 178 & 178 \\ \hline 
Masses & & & \\ \hline
$|\mu |$       & 569 &1186  &1171  \\ \hline
h$^0$          & 119 & 124  & 124  \\
H$^0$          & 626 &1293  &1261  \\
A$^0$          & 622 &1285  &1253  \\
H$^{\pm}$      & 632 &1296  &1264  \\ \hline
$\chi^0_1$     & 175 & 417  & 417  \\
$\chi^0_2$     & 339 & 805  & 804  \\
$\chi^0_3$     & 574 &1192  &1176  \\
$\chi^0_4$     & 587 &1200  &1184  \\
$\chi^{\pm}_1$ & 340 & 807  & 806  \\
$\chi^{\pm}_2$ & 587 &1200  &1184  \\ \hline
$\tilde{g}$    &1026 &2191  &2191  \\ \hline
$e_L$, $\mu_L$ & 306 & 684  & 677  \\
$e_R$, $\mu_R$ & 171 & 387  & 374  \\
$\nu_e$, $\nu_{\mu}$
               & 290 & 669  & 662  \\
$\tau_1$       & 153 & 338  & 319  \\
$\tau_2$       & 309 & 677  & 670  \\
$\nu_{\tau}$   & 288 & 660  & 653  \\ \hline
$u_L$, $c_L$   & 935 &1991  &1988  \\
$u_R$, $c_R$   & 902 &1911  &1908  \\
$d_L$, $s_L$   & 938 &1993  &1990  \\
$d_R$, $s_R$   & 899 &1903  &1900  \\
$t_1$          & 710 &1545  &1553  \\
$t_2$          & 900 &1842  &1840  \\ 
$b_1$          & 852 &1807  &1804  \\
$b_2$          & 883 &1851  &1846  \\ 
\hline
\end{tabular}
\caption{Proposed GDM benchmark points taken from \cite{DeRoeck:2005bw}. Mass
spectra [GeV] are calculated using {\tt
ISAJET~7.69}~\protect\cite{Baer:1999sp}.}
\label{table:benchmarks}}
\end{TABLE}

%\pagebreak

One of the most important decays of these GDM models, at least for
those within the reach of the LHC, is that of the right-handed squark:
$\tilde q_R\to q\tilde\chi_1^0$ with branching ratios of almost 100\%
at all three points. This is followed by the decay
$\tilde\chi_1^0\to\tau\tilde\tau_1$ in a large fraction of events
(92/75/69 \% for the three points), so that the decays of two
right-handed squarks result in the relatively clean final states
$qq\tau\tau\tilde\tau_1\tilde\tau_1$ in a large fraction of the
events. From the branching ratio of this decay chain shown in the left
plot of Fig.~\ref{fig:scan1}, we see that this signature is a quite
generic feature for the class of GDM models with a $\tilde\tau$
NLSP. In the experiments it should be possible to trigger on the pair
of high-$p_T$ jets produced in the primary $\tilde q_R\to
q\tilde\chi_1^0$ or the pair of tau jets, and one can then identify
the stau by its peculiar signature as a slow moving muon and measure
its mass and momentum to high precision, as discussed in the next
Section. One can also identify the decay products of the $\tau$ and
reconstruct the full $\tau$ momenta using transverse momentum balance,
as we show in Section~\ref{sect:fastsim}. Thus, these events can be
fully reconstructed and we can make a precise determination of the
masses of all three supersymmetric particles involved in the decay
cascade. From this decay chain one may also be able think of
reconstructing the gluino $\tilde g$ from its decays into squark-quark
pairs, as we discuss later for benchmark point $\epsilon$.

%%%%%%%%%%%%%%%%%%%%%%%%%%%%%%%%%%%%%%%%%%%%%%%%%%%%%%%%%%%%%%%%%%%%%%%%
\FIGURE[htb]{
\epsfxsize 15cm
\epsfbox{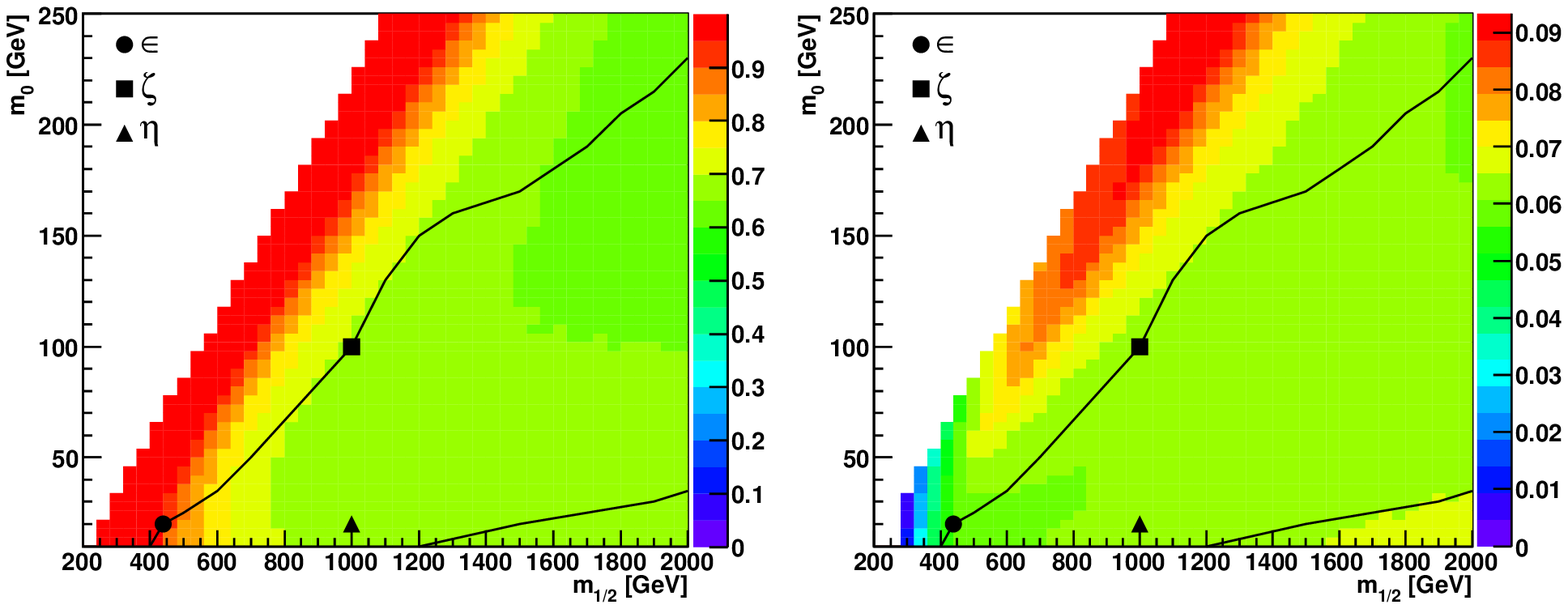}
\caption{Total branching ratios for the
$\tilde q_R\to q\tilde\chi_1^0\to q\tau\tilde\tau_1$ (left) and
$\tilde q_L\to q\tilde\chi_2^0\to q\ell\tilde\ell_L\to
q\ell\ell\tilde\chi_1^0\to q\ell\ell\tau\tilde\tau_1$ (right) decay
chains in the $(m_{1/2}, m_0)$ parameter plane. The solid black lines
indicate the boundary of the cosmologically allowed region of
parameter space explored in~\protect\cite{DeRoeck:2005bw}.}
\label{fig:scan1}
}
%%%%%%%%%%%%%%%%%%%%%%%%%%%%%%%%%%%%%%%%%%%%%%%%%%%%%%%%%%%%%%%%%%%%%%

%%%%%%%%%%%%%%%%%%%%%%%%%%%%%%%%%%%%%%%%%%%%%%%%%%%%%%%%%%%%%%%%%%%%%%%%
\FIGURE[htb]{
\epsfxsize 15cm
\epsfbox{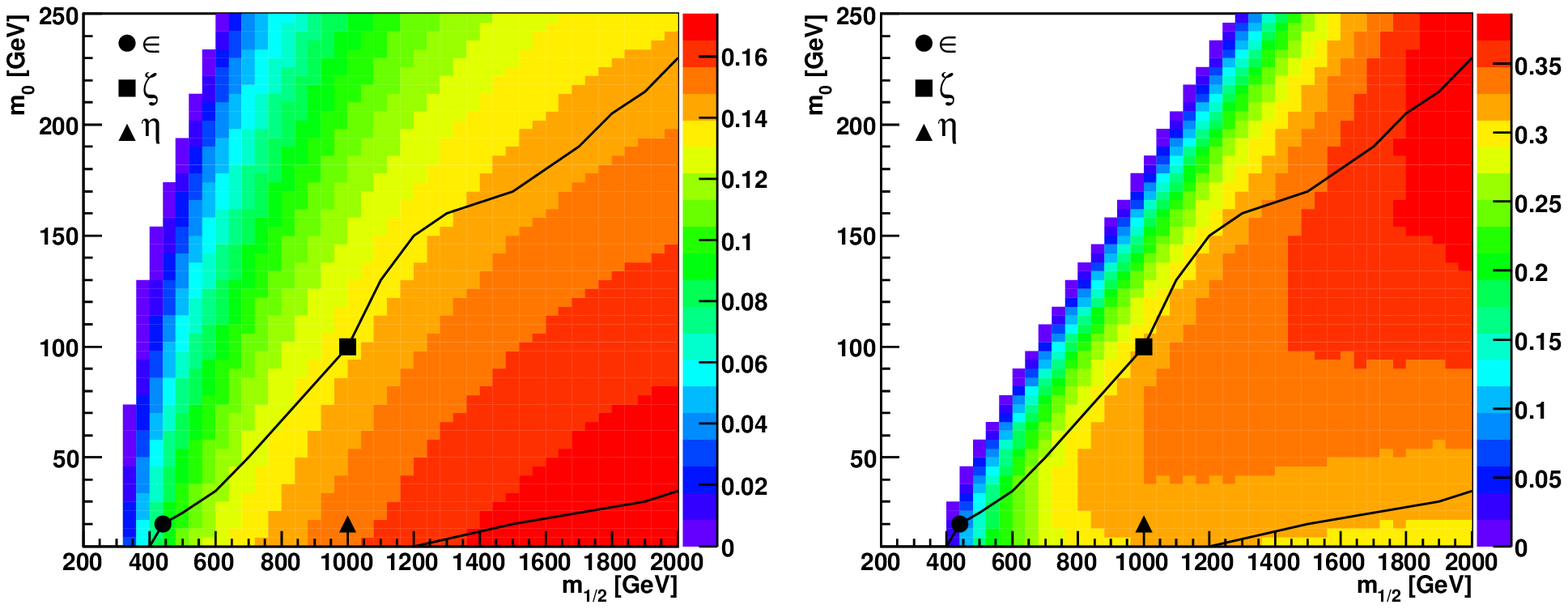}
\caption{Branching ratio for the decay chains
$\tilde\chi_1^\pm\to\tau\tilde\nu_\tau\to\tau W\tilde\tau_1$ (left)
and $\tilde\chi_1^0\to \ell\tilde\ell_R\to\ell\ell\tau\tilde\tau_1$
(right) in the $(m_{1/2}, m_0)$ parameter plane. The solid black lines
indicate the cosmologically allowed region of parameter space explored
in~\protect\cite{DeRoeck:2005bw}.}
\label{fig:scan2}
}
%%%%%%%%%%%%%%%%%%%%%%%%%%%%%%%%%%%%%%%%%%%%%%%%%%%%%%%%%%%%%%%%%%%%%%

The decays of left-handed squarks have in general more leptons in the
final state, for example when decaying via
$\tilde\chi_2^0\to\ell\tilde\ell_L$ and
$\tilde\ell_L\to\ell\tilde\chi_1^0$, with the total branching ratio
shown in the right plot of Fig.~\ref{fig:scan1}. With knowledge of the
$\tilde\chi_1^0$ mass, this makes possible the reconstruction of the
left-handed slepton and the $\tilde\chi_2^0$. However, left-handed
squark decays into charginos are more difficult to use because of
neutrinos in the final state, either directly from the chargino decay,
or from the leptonic decay of a $W$.

Whilst the branching ratio of the left-handed squark is too low to
search for events with two such decay chains, the considerations for
triggering, particle identification and reconstruction are similar to
those for the right-handed squark. The appearance of hard leptons help
identify the decay chain, but give additional combinatorial
difficulties in particle identification as we will see in
Section~\ref{sect:fastsim}. Nevertheless, it seems possible to
reconstruct this decay chain, at least for the benchmark point
$\epsilon$, as we discuss later.

At the $\epsilon$ benchmark point the supersymmetric events are
dominated by gluino/squark production followed by cascade decays. For
the low cross-section benchmark points $\zeta$ and $\eta$, we see from
Table~\ref{table:xsec} that these events make up a smaller fraction of
the total number of supersymmetric events. Here Drell-Yan and
associated neutralino and chargino production become more
important. These events feature less jet activity and heavier, slower
staus, and may lead to some trigger problems (see
Section~\ref{sect:trigger}). By considering only the Drell-Yan
cross-section, taking into consideration triggering and reconstruction
efficiencies for stau pair production, one could investigate the reach
of the LHC in setting model-independent exclusion limits on the mass
of the stau. However, this would require a very detailed understanding
of the effects of the background cuts suggested in
Section~\ref{sect:background} on SM events in the limit where the
background tends to zero, and is perhaps best left to a full detector
simulation. Here we will instead focus on some interesting decay
chains starting from the associated production. The branching ratio of
the decay $\tilde\chi_1^\pm\to\tau\tilde\nu_\tau\to\tau W\tilde\tau_1$
is shown in Fig.~\ref{fig:scan2}, and is almost identical to that for the
decay $\tilde\chi_2^0\to\nu_\tau\tilde\nu_\tau\to\nu_\tau
W\tilde\tau_1$, as the decays differ essentially only by the mass of
the tau. If we can reconstruct the $W$ produced in both cases from its
hadronic decay and combine with the correct stau candidate, this gives
us direct access to the tau-sneutrino and, possibly, chargino masses.

Among the lighter particles of Table~\ref{table:benchmarks} that one would
expect to be copiously produced at the LHC in the case the $\epsilon$
benchmark, we have not yet discussed decays involving the $\tilde\ell_R$,
the right-handed selectron or smuon. These are dominantly
produced in decays of neutralinos, the only exception being the
direct pair-production of the sleptons, which have low cross sections. In
fact, only in the decay of the lightest neutralino is there a
significant rate, due to the preference for decays to left-handed
sleptons, and even here it is suppressed relative to decays to the lighter
stau. The $\tilde\ell_R$ decays in turn almost exclusively to the three
bodies $\ell\tau\tilde\tau_1$. In the right plot of
Fig.~\ref{fig:scan2} we show the total branching ratio of the decay
$\tilde\chi_1^0\to \ell\tilde\ell_R\to\ell\ell\tau\tilde\tau_1$. One
could hope to reconstruct the $\tilde\ell_R$ mass from this decay,
starting from the stau. However, the low branching ratio at the point
$\epsilon$, combined with the soft spectra of the leptons from the
decay, due to the small mass differences, makes this very
difficult. Our attempts leave us with only a few events. The larger
branching ratios and harder leptons of the two other benchmarks are
unfortunately balanced out by the overall much lower cross sections. It
thus seems to be very challenging to reconstruct the $\tilde\ell_R$ in
GDM models at the LHC, and we do not discuss the $\tilde\ell_R$
further.

The first four decay chains we discussed above will be the main focus
of our attempts at reconstructing supersymmetric masses in
Section~\ref{sect:fastsim}. The fact that, as shown in
Figs.~\ref{fig:scan1}~and~\ref{fig:scan2}, the total branching ratios
for these decay chains are relatively high across large regions of the
$(m_{1/2}, m_0)$ plane, with a dominating branching ratio for the
chain initiated by $\tilde q_R\to q\tilde\chi_1^0$ in all the explored
range of the allowed parameter space, up to values of $m_{1/2}\sim
2$~TeV, suggests that the techniques discussed here would be of wide
utility at the LHC.

For the most part, we limit our discussions to simulations of the
high-cross-section benchmark point $\epsilon$, and only summarise our
results for the other two points. We draw attention to instances where
there are important differences between the benchmark points, and we
refer the reader to~\cite{DeRoeck:2005bw} for further details about
them.

%%%%%%%%%%%%%%%%%%%%%%%%%%%%%%%%%%%%%%%%%%%%%%%%%%%%%%%%%%%%%%%%%%%%%%
\section{Fast Monte Carlo Simulation}
\label{sect:MC}
%%%%%%%%%%%%%%%%%%%%%%%%%%%%%%%%%%%%%%%%%%%%%%%%%%%%%%%%%%%%%%%%%%%%%%

For the fast simulation we have generated proton-proton collisions at
the LHC energy using {\tt PYTHIA\,6.326} \cite{PYTHIA} and CTEQ 5M1
parton distribution functions \cite{Lai:1999wy}. For the $\epsilon$
benchmark point we show results on mass measurements for a total
integrated luminosity of $30~\text{fb}^{-1}$, corresponding to the
planned initial running of the LHC at low luminosity. In the cases of
the $\zeta$ and $\eta$ points, which have relatively low cross
sections, we show results for statistics equivalent to
$300~\text{fb}^{-1}$. These may then be viewed as estimates of the
ultimate precisions achievable for these benchmark points without an
upgrade of the LHC luminosity.

%%%%%%%%%%%%%%%%%%%%%%%%%%%%%%%%%%%%%%%%%%%%%%%%%%%%%%%%%%%%%%%%%%%%%%
\subsection{Parametrisation of Detector}
\label{sect:parametrization}
%%%%%%%%%%%%%%%%%%%%%%%%%%%%%%%%%%%%%%%%%%%%%%%%%%%%%%%%%%%%%%%%%%%%%%

The detector simulation has been carried out using the generic LHC
detector simulation {\tt AcerDET\,1.0}~\cite{Richter-Was:2002ch},
taking the ATLAS detector as our model for a LHC detector. Here we
give a short summary of the most important choices made for the {\tt
AcerDET} settings: We consider a lepton to be identified if it has
transverse momentum $p_T>5(6)$~GeV and $|\eta|<2.5$ for electrons
(muons). A lepton is considered isolated if it is at a distance
$\Delta R>0.4$, where $\Delta R \equiv
\sqrt{(\Delta\phi)^2+(\Delta\eta)^2}$, from other leptons and jets,
and if the transverse energy deposited in a cone $\Delta R=0.2$ around
the lepton is less than $10$~GeV. Jets are reconstructed by a
cone-based algorithm from clusters and are accepted if the jet has
$p_T>15$~GeV within a cone $\Delta R=0.4$. The jets are re-calibrated
using an included flavour-independent parametrisation, optimised to
give the correct resonance mass in dijet decays. We use the $p_T$
parametrised $b$-tagging efficiency and light jet rejection, for a
low-luminosity environment, given in~\cite{unknown:1997fs}. For tau
jets we use a lower cut on transverse momentum of $p_T>20$~GeV, and
the parametrised tau-tagging efficiency and rejection factors
from~\cite{heldmann}.

For the detector's response to staus, we take our parametrisation of
the momentum resolution from~\cite{polesello,Ambrosanio:2000ik}, which
performed a full simulation of the response of the ATLAS muon system
to a metastable stau of mass $101$~GeV, in a Gauge Mediated
Supersymmetry Breaking (GMSB) model. This momentum dependent
resolution is used to smear the stau momentum in the detector
simulation. We have performed a smearing with width $\sigma_p$, given
by
\begin{equation}
\frac{\sigma_p}{p} = k_1 p \oplus k_2\sqrt{1+\frac{m_{\tilde\tau_1}^2}{p^2}}
\oplus \frac{k_3}{p},
\end{equation}
where $k_1$ is the parameter of the sagitta measurement error, $k_2$
represents the multiple scattering and $k_3$ the fluctuation of energy
loss in the calorimeter. From~\cite{polesello} we take the values
$k_1=0.0118$, $k_2=0.02$ and $k_3=89$.

For the measurement of the velocity of staus and muons, we use the
parametrisation of the velocity resolution
from~\cite{polesello,Ambrosanio:2000ik}. In our simulation we smear
the true velocity by:
\begin{equation}
\frac{\sigma_\beta}{\beta}=0.028\,\beta,
\label{eq:sigmabeta}
\end{equation}
which was found by looking at the fit quality of tracks in the muon
system as a function of the particle's assumed arrival time in the
Monitored Drift Tubes (MDT), subsequently minimising the $\chi^2$ as a
function of the arrival time to find the correct Time-of-Flight
(ToF). This is a conservative estimate of the resolution as the
simulation was carried out in the central detector, for
$\eta=0.1$. For larger pseudo-rapidities the resolution is expected to
improve, due to a longer flight path. An alternative way to measure
the ToF would be to use the timing of hits in the Resistive Plate
Chambers (RPC), responsible for muon triggering, which may be able to
provide similar precision.

%%%%%%%%%%%%%%%%%%%%%%%%%%%%%%%%%%%%%%%%%%%%%%%%%%%%%%%%%%%%%%%%%%%%%%
\subsection{Trigger}
\label{sect:trigger}
%%%%%%%%%%%%%%%%%%%%%%%%%%%%%%%%%%%%%%%%%%%%%%%%%%%%%%%%%%%%%%%%%%%%%%

It was suggested in~\cite{DeRoeck:2005bw} that the triggering
efficiency for events in GDM scenarios will be high. A full simulation
of these benchmark points, which will include triggers, is under
way~\cite{fullsim}, and we have every reason to believe that due to
the generically large jet and lepton activity in the supersymmetric
events we will have trigger efficiencies in the high $90$\% range for
all three benchmarks. As an estimate of the trigger, in our fast
simulation we have required each event to have at least one of the
following:
\begin{itemize}
\item
A jet with $p_T>290$~GeV,
\item
three jets with $p_T>130$~GeV,
\item
four jets with $p_T>90$~GeV,
\item
one muon or stau with $p_T>20$~GeV, or
\item
two muons or staus with $p_T>6$~GeV.
\end{itemize}
These are the jet and muon trigger thresholds for the ATLAS detector,
taken from~\cite{:1999fq}, where we conservatively use the 
high-luminosity thresholds and ignore electromagnetic, missing-energy and
tau triggers. Since one may fail to trigger on a slow stau in the
normal running of the LHC detectors because it arrives at the trigger
station too late, we additionally require triggering staus to have a
velocity of $\beta\gamma>0.9$. This number is a conservative estimate
from the ATLAS geometry, to give a trigger in the correct bunch
crossing. Staus with $|\eta|\approx 1$ have the longest distance to
travel between RPC layers. Given a $25$~ns gap between bunch crossings
and a distance of $\approx 15.5$~m to the outer RPC layer the critical
velocity is $\beta=0.67$, or $\beta\gamma=0.9$, ignoring any
significant energy loss. 
%The resulting trigger efficiencies is $98.5$\%
%for $\epsilon$ and $97.7$\% for both $\zeta$ and $\eta$.

%%%%%%%%%%%%%%%%%%%%%%%%%%%%%%%%%%%%%%%%%%%%%%%%%%%%%%%%%%%%%%%%%%%%%%
\subsection{Background}
\label{sect:background}
%%%%%%%%%%%%%%%%%%%%%%%%%%%%%%%%%%%%%%%%%%%%%%%%%%%%%%%%%%%%%%%%%%%%%%

In order to separate the staus from background muons in the supersymmetric
events, and to remove the Standard Model background, we use the following cuts,
which we will refer to as the standard GDM cuts:
\begin{itemize}
\item
There should be exactly two stau candidates in each event,
\item
these should be isolated and within the geometrical acceptance of the
muon chambers as per the requirements for the muons listed above,
\item
they should have $p_T>50$~GeV,
\item
they should have a velocity $\beta\gamma<6.0$, and
\item
they should have a mass estimate from their momentum and velocity that
is consistent with the stau mass after it has been measured (see
Section~\ref{sect:mstau});
\item 
in addition, the sum of the number of leptons (muons or electrons)
with $p_T>10$~GeV and tau-tagged jets\footnote{Assuming a $50$\% tau
tagging efficiency.} with $p_T>20$~GeV in each event should be exactly
two.
\end{itemize}
The final cut is made on the basis that in GDM scenarios with a
$\tilde\tau_1$ NLSP, the vast majority of events will contain two taus
in addition to the pair of staus, and these should be identified to
fully reconstruct the particles in the supersymmetric decay chains
(see Section~\ref{sect:tau}). Note that we have a much looser cut on
$\beta\gamma$ than used in~\cite{Ambrosanio:2000ik}. We still expect
to remove most SM background because of the requirement of two stau
candidates in each event, and in particular because of the extra
requirement of two tau candidates.

After our trigger requirements and these cuts, we should have a very
powerful rejection on Standard Model background events. We have tested
this hypothesis in a fast simulation of a background sample consisting
of the equivalent of $30~\text{fb}^{-1}$ of $t\bar t$ events with an
NLO cross section of $737$~pb~\cite{Frixione:2003ei}, and samples of
$p_T$ binned QCD, $W$+jets, $Z$+jets and $WW/WZ/ZZ$ production
events. The total numbers of events are $1.75\cdot 10^6$ for QCD and
$2.5\cdot 10^5$ for the other processes. We find no events that
survive the cuts listed above, and, in order to find the first events
that pass the cuts, we would need to relax the velocity cut to
$\beta\gamma<8.3$, after which we find one $WW/WZ/ZZ$ event that
passes. However, with the limited statistics available, in particular
for QCD, this null result must be interpreted with care. For example,
the parametrisation of the $\beta$ resolution assumes a Gaussian
shape. If the true distribution has significantly larger tails, this
will affect the power of the cut on $\beta\gamma$. A full simulation
study of the muon velocity resolution would be required to investigate
this possibility. It is partially because of this issue that we prefer
to use a loose cut on $\beta\gamma$. Despite these problems, with
the additional cuts made in the searches for various supersymmetric
particles presented in Section~\ref{sect:fastsim}, we expect that the
cuts presented here will render negligible the Standard Model
background in our analysis.

%%%%%%%%%%%%%%%%%%%%%%%%%%%%%%%%%%%%%%%%%%%%%%%%%%%%%%%%%%%%%%%%%%%%%%
\section{Reconstruction of Masses in GDM Models}\label{sect:fastsim}
%%%%%%%%%%%%%%%%%%%%%%%%%%%%%%%%%%%%%%%%%%%%%%%%%%%%%%%%%%%%%%%%%%%%%%
In this Section, we incorporate the parametrisation of the response of
the ATLAS detector to the slow-moving $\tilde\tau_1$s from
Section~\ref{sect:MC} in a fast simulation of sparticle pair
production and cascade decays at an LHC detector. This enables us to
assess the possibility of identifying various heavier sparticles and
the precision obtainable in mass measurements in GDM models. Since we
have no escaping LSP in these GDM models, we are not limited to the
standard measurements of the endpoints and shapes of kinematic
distributions, see
e.g.~\cite{Hinchliffe:1996iu,Bachacou:1999zb,Allanach:2000kt,
Lester:2001zx,Gjelsten:2004,Weiglein:2004hn,Gjelsten:2005aw,Miller:2005zp},
and we can fully reconstruct sparticles, starting from the stau NLSP.

We begin in Section~\ref{sect:mstau} by looking at the starting point
of all our mass measurements, the measurement of the stau mass. Then,
in Section~\ref{sect:tau} we demonstrate the complete reconstruction
of the tau momenta for events with only two taus and no hard
neutrinos. Section~\ref{sect:qRchain} deals with the reconstruction of
the decay chain $\tilde q_R \to q
\tilde\chi_1^0 \to q\tau\tilde\tau_1$, and Section~\ref{sect:qLchain}
deals with the reconstruction of the cascade $\tilde q_L \to q
\tilde\chi_2^0 \to q \ell\tilde\ell \to q \ell\ell\tilde\chi_1^0
\to q \ell\ell\tau\tilde\tau_1$.
Finally, in Section~\ref{sect:associated} we discuss the possibility of
measuring sneutrino and chargino masses using dijet decays of $W$
bosons produced in supersymmetric cascade decays.

%%%%%%%%%%%%%%%%%%%%%%%%%%%%%%%%%%%%%%%%%%%%%%%%%%%%%%%%%%%%%%%%%%%%%%
\subsection{Measuring the Stau Mass}\label{sect:mstau}
%%%%%%%%%%%%%%%%%%%%%%%%%%%%%%%%%%%%%%%%%%%%%%%%%%%%%%%%%%%%%%%%%%%%%%

We first present an estimate of the achievable precision on the
measurement of the stau mass for the parametrisation of the momentum
and velocity resolution given in Section~\ref{sect:parametrization}.
After the trigger simulation presented in Section~\ref{sect:trigger},
and making the standard GDM cuts of Section~\ref{sect:background} with the
exception of the cut on the stau mass and the requirement of two tau
candidates, we plot in Fig.~\ref{fig:mstau} (left) a scatter plot of
the measured velocity $\beta\gamma_{\rm{meas}}$ versus the measured
mass for single staus, as inferred from
\begin{equation}
m_{\tilde\tau_1} = \frac{p_{\rm{meas}}}{\beta\gamma_{\rm{meas}}}.
\end{equation}

%%%%%%%%%%%%%%%%%%%%%%%%%%%%%%%%%%%%%%%%%%%%%%%%%%%%%%%%%%%%%%%%%%%%%%%%
\FIGURE[htb]{
\epsfxsize 15cm
\epsfbox{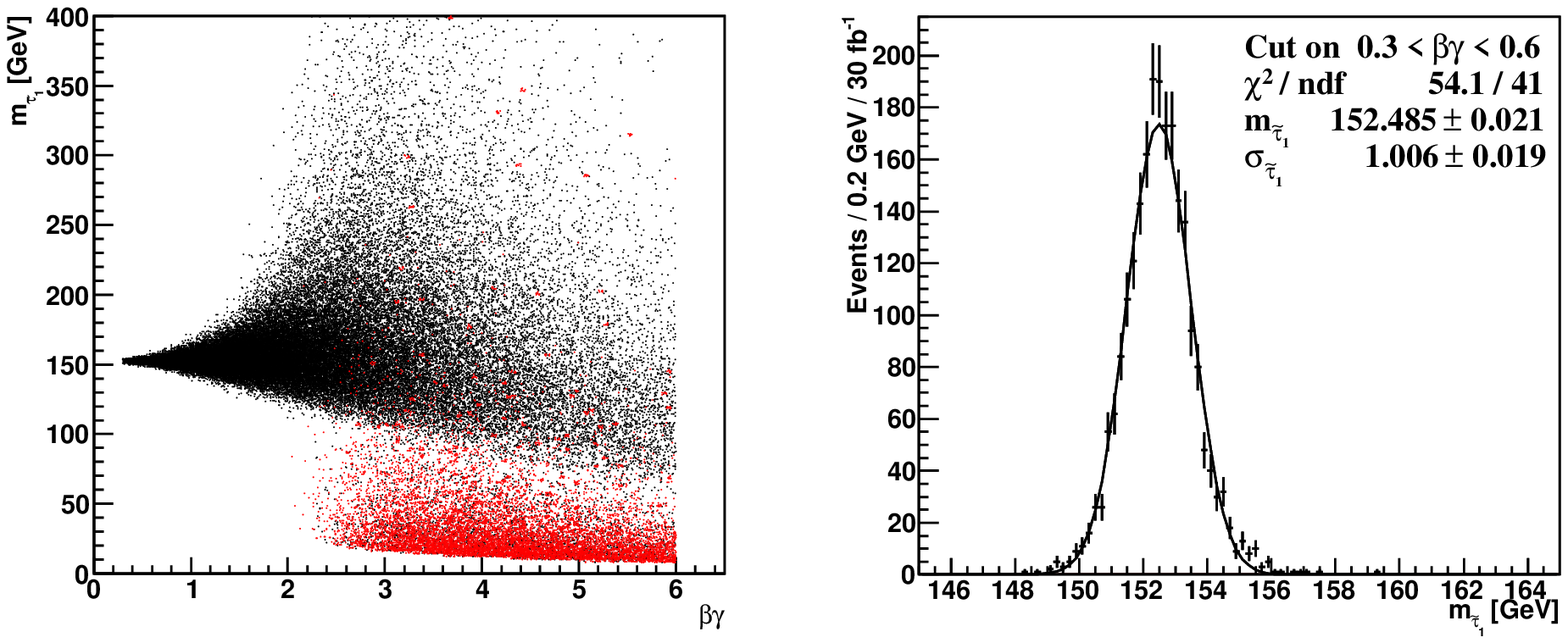}
\caption{Scatter plot of measured velocity $\beta\gamma_{\rm{meas}}$
versus measured mass (left), with supersymmetric events in black and
SM background events in red, and a corresponding plot of the measured
stau mass (right) with an additional cut on the velocity of
$0.3<\beta\gamma<0.6$.}
\label{fig:mstau}
}
%%%%%%%%%%%%%%%%%%%%%%%%%%%%%%%%%%%%%%%%%%%%%%%%%%%%%%%%%%%%%%%%%%%%%%

\noindent
We see a large spread in the masses for high $\beta\gamma$,
corresponding to high momenta. This is expected as the assumed
momentum resolution deteriorates significantly above a few hundred
GeV, and the velocity resolution worsens at higher velocity due to the
shorter ToF. One effect of the momentum dependence of the resolution
is that the spread has a clear bias towards higher masses, which would
be a systematic effect for a mass measurement using all events passing
the cuts. We also see a grouping of events at high $\beta\gamma$ and
low masses. These are mis-measured muons that have been assigned velocities that are too low,
and thus passed the GDM cuts used. This is particularly clear
for the SM background events which are shown in red.

Because the low-velocity events have a much smaller spread and less
bias towards higher masses, it would be advantageous to use only these
in a mass measurement. As we commented in Section~\ref{sect:trigger},
the staus from these events will be missed by the trigger system
because they are too slow. This does not mean that they would be
missed completely by the LHC detectors, since their muon systems are
designed sufficiently robustly as to record hits in a large time
window after a collision. However, it does mean that we will have to
rely on other triggers, e.g., calorimetric triggers, or triggers on
muons from the decay of the heavier sparticles. Exactly how far down
in velocity it is possible to go must be the subject of full detector
studies that lie beyond the scope of this paper. In~\cite{polesello}
the velocities of the staus were found to be measurable at least down
to $\beta\gamma = 0.44$, albeit for a simulation of single staus in
the muon system only. However, in~\cite{Ambrosanio:2000ik} a conservative
approach to the trigger issue was followed, ignoring events where the
staus did not reach the muon chambers inside the time window of the
muon triggers, and using a lower cut of $\beta\gamma \ge 0.75$. Assuming
that the trigger rates from our simplified trigger cuts are not
entirely unrealistic and that we can reconstruct staus with velocities
down to $\beta\gamma\approx 0.3$, we make an additional cut on
$\beta\gamma<0.6$ for the sample used to measure the stau mass, and
plot the corresponding mass distribution in Fig.~\ref{fig:mstau}
(right). Fitting the mass distribution with a Gaussian we get a stau
mass for the $\epsilon$ benchmark of
\begin{equation}
m_{\tilde\tau_1} \; = \; 152.485 \pm 0.021~{\rm GeV,}
\label{eq:mstau}
\end{equation}
to be compared with the nominal value of
$m_{\tilde\tau_1}=152.475$~GeV.

To investigate the dependence of the measured stau mass on the lower
limit on the velocity of staus that we can identify, we show in
Fig.~\ref{fig:mstaubg} the reconstructed stau mass as a function of
the lower $\beta\gamma$ cut value for three different values of the
upper cut on $\beta\gamma$. The general tendency is, as could be
expected, an increase in the systematic error, represented by the
increase in distance from the nominal mass value, as the lower cut
value is raised and low-velocity events are discarded~\footnote{It is
possible that this systematic effect could be modelled and (at least
partially) mitigated in a more complete analysis.}. As the lower cut
approaches the upper cut value, this effect flattens out, while of
course the statistical error increases with the reduction in the
number of events. It is clear that, for a lower cut on $\beta\gamma$,
in choosing the upper cut one must trade the loss of statistics
against the reduction of systematic error from the higher-velocity
staus. We see that there is very little change in the stau mass 
determination if the lower limit on $\beta\gamma$ is increased to 
0.44~\cite{polesello}. We also note that the error in $m_{\tilde \tau}$
remains below the per-mille level even if we restrict the analysis to
$0.6 < \beta\gamma < 0.9$. In all the examples shown in Fig.~\ref{fig:mstaubg}, the error
in calculating $m_{\tilde \tau}$ is negligible compared with the
errors encountered in the reconstruction of higher-mass states
decaying into staus.

%%%%%%%%%%%%%%%%%%%%%%%%%%%%%%%%%%%%%%%%%%%%%%%%%%%%%%%%%%%%%%%%%%%%%%%%
\FIGURE[htb]{
\epsfxsize 7.5cm
\epsfbox{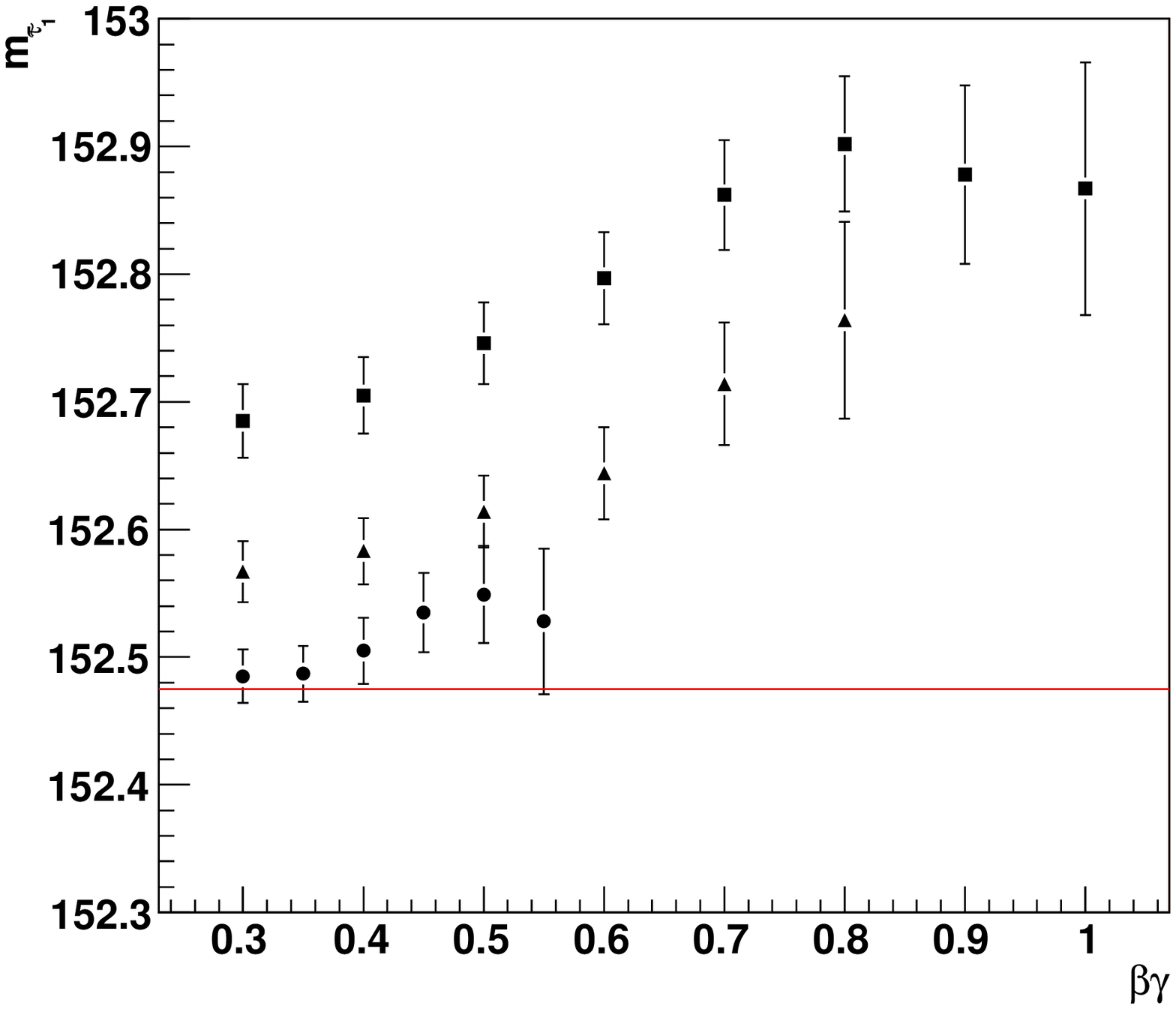}
\caption{The stau mass from a Gaussian fit as a function of a lower cut
on $\beta\gamma$. The solid circles show the results for an upper cut
of $\beta\gamma<0.6$, the triangles $\beta\gamma<0.9$ and the squares
$\beta\gamma<1.2$. The red line shows the nominal mass value.}
\label{fig:mstaubg}
}
%%%%%%%%%%%%%%%%%%%%%%%%%%%%%%%%%%%%%%%%%%%%%%%%%%%%%%%%%%%%%%%%%%%%%%

Based on the clear separation of muons and staus in the scatter plot
of Fig.~\ref{fig:mstau}, we require in the further analyses that stau
candidates have inferred masses above $100$~GeV. This value was also
used in the background analysis of Section~\ref{sect:background}. We
use the stau mass of (\ref{eq:mstau}) in the following discussions,
although the much larger errors involved in the reconstruction of
heavier sparticles means that we are insensitive to the exact value,
within the errors discussed in this Section. The results for the other
two benchmarks are analogous, and the resulting masses can be found in
Table~\ref{table:mm}. The statistical errors are at the sub-permille
level, which is below the expected uncertainty on the lepton energy
scale of order $0.1$\%, and on the same level as the results given for
various GMSB scenarios in~\cite{Ambrosanio:2000ik}. From
Fig.~\ref{fig:mstaubg}, we conclude that it should be possible also to
keep the systematic error from the momentum and velocity measurements
at the same level if we are able to reconstruct staus with velocities
down to $\beta\gamma=0.6$.

The possibility of making further measurements on the staus by looking
for staus trapped in the LHC detectors or in surrounding matter was
discussed in~\cite{DeRoeck:2005bw}. Trapped, decaying staus could make
a measurement of the stau lifetime possible, leading to an indirect
determination of the gravitino mass, assuming the macroscopically
determined value of the Planck scale, and could potentially even lead
to a microscopic determination of the Planck scale and a test of
supergravity if the gravitino mass could be measured directly from the
decay kinematics~\cite{Weiglein:2004hn,Buchmuller:2004rq}. However,
these exciting possibilities lie somewhat outside of the scope of this
paper.

\begin{TABLE}{
\begin{tabular}{|c||r|r|r|}
\hline
Model            & $\epsilon$~~~~~~~~~& $\zeta$~~~~~~~   & $\eta$~~~~~~~    \\ \hline
Mass             &                    &                  &                  \\
$\tilde\tau_1$   & $152.485\pm 0.021$ & $338.24\pm 0.09$ & $319.02\pm 0.09$ \\
$\tilde\nu_\tau$ & $291.8\pm 1.2$     & $666.4\pm 1.9$   & $659.6\pm 1.1$   \\
$\tilde\ell_L$   & $307.9\pm 2.0$     & $674.6\pm 3.4$   & $673.5\pm 3.5$   \\
$\tilde\chi^0_1$ & $176.6\pm 0.9$     & $418.3\pm 1.8$   & $419.6\pm 3.7$   \\
$\tilde\chi^0_2$ & $339.2\pm 2.0$     & $(796.5\pm 6.5)$ & $(801.0\pm 7.2)$ \\
$\tilde q_L$     & $923.4\pm 3.7$     & -                & -                \\
$\tilde q_R$     & $895.8\pm 2.8$     & $1811.2\pm 332.3$& -                \\
$\tilde b$       & $888.9\pm 57.7$    &  -               & -                \\
\hline
\end{tabular}
\caption{The mass measurements and statistical errors [in GeV] obtained
in the simulations of the GDM benchmark points. The numbers in
brackets (~) are obtained from a fit to the edge in the dilepton
spectrum. Jet and lepton energy scale errors are not included.}
\label{table:mm}}
\end{TABLE}

%%%%%%%%%%%%%%%%%%%%%%%%%%%%%%%%%%%%%%%%%%%%%%%%%%%%%%%%%%%%%%%%%%%%%%
\subsection{Recalibrating Taus}\label{sect:tau}
%%%%%%%%%%%%%%%%%%%%%%%%%%%%%%%%%%%%%%%%%%%%%%%%%%%%%%%%%%%%%%%%%%%%%%

%%%%%%%%%%%%%%%%%%%%%%%%%%%%%%%%%%%%%%%%%%%%%%%%%%%%%%%%%%%%%%%%%%%%%%%%
\FIGURE[htb]{
\epsfxsize 7.5cm
\epsfbox{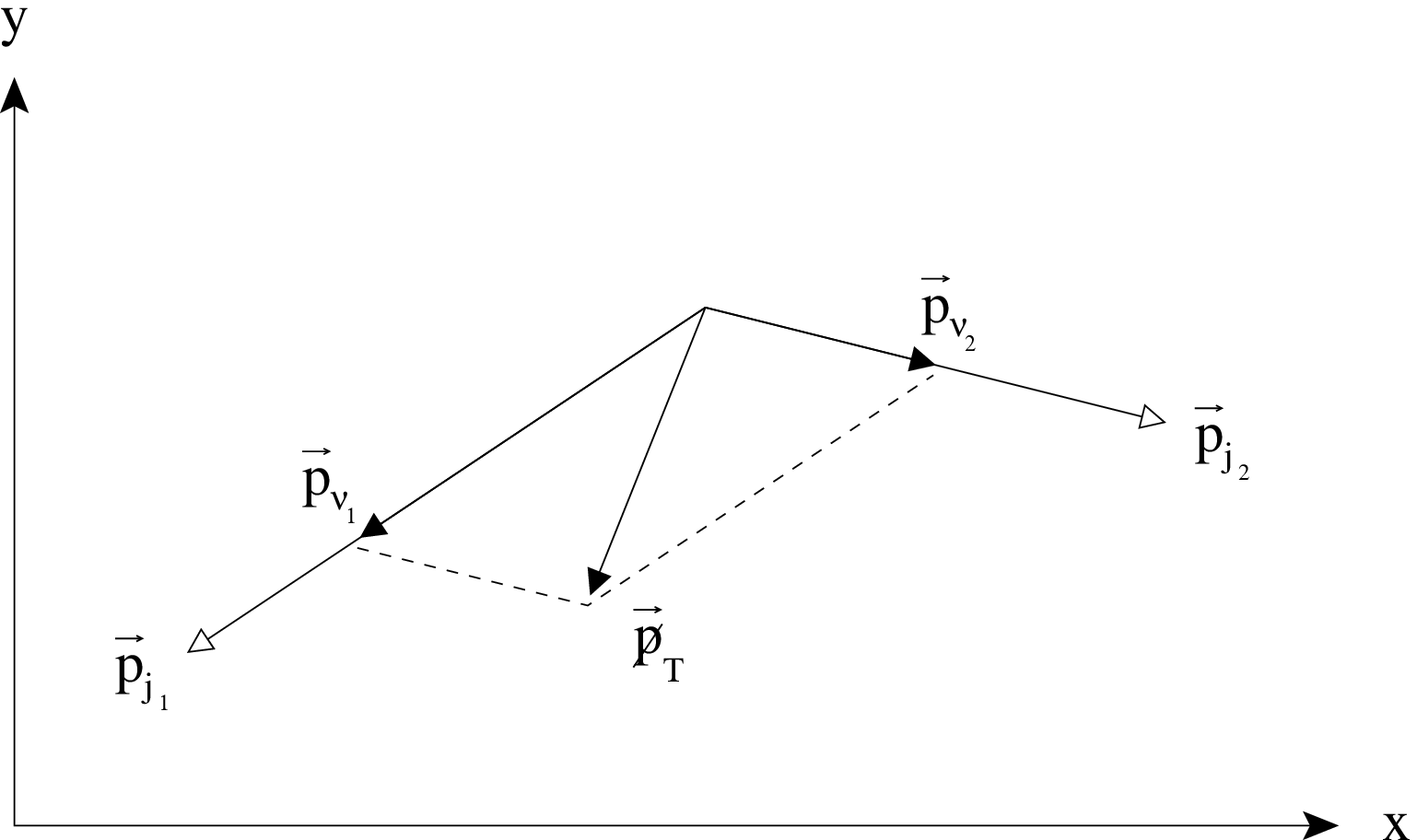}
\caption{Schematic drawing of the $\tau$ decay products
$\vec p_{j_1}$ and $\vec p_{j_2}$, the missing momentum $\not\!\!\vec
p_{T}$ and the reconstructed neutrino momenta $\vec p_{\nu_1}$ and
$\vec p_{\nu_2}$, projected on the transverse plane.}
\label{fig:rec}
}
%%%%%%%%%%%%%%%%%%%%%%%%%%%%%%%%%%%%%%%%%%%%%%%%%%%%%%%%%%%%%%%%%%%%%%

As was already mentioned in Section~\ref{sect:background}, in all
R-parity conserving GDM scenarios with a $\tilde \tau_1$ NLSP, with
the exception of Drell-Yan pair productions of staus and some events
yielding tau neutrinos, each event will contain two $\tau$s in
addition to the pair of $\tilde \tau_1$s. In Section~\ref{sect:mstau}
we discussed the measurement of the $\tilde\tau_1$ mass. We can detect
and measure the masses of many of the heavier sparticles by
reconstructing decay chains that lead to the $\tilde\tau_1$s. For
example, at benchmark point $\epsilon$, essentially $100$\% of the
lightest neutralinos ${\tilde\chi_1^0}$ decay into $\tilde\tau_1 +
\tau$, and this branching ratio is also very large at the other two
points ($\sim 70$\%). One can reconstruct a ${\tilde\chi_1^0}$ mass
peak by combining $\tilde\tau_1$-$\tau$ pairs, but to do this one
needs to know the momenta of the accompanying $\tau$s, which in turn
decay and lose momentum to escaping neutrinos.

Assuming that a $\tau$ is relativistic in the laboratory frame, as
will indeed be the case for any $\tau$ that would pass our acceptance
criteria, its hadronic decay will give a jet and a neutrino travelling
in essentially the same direction. If there are exactly two $\tau$s in
an event, and no other source of missing momentum, the sum of the
missing transverse momenta from the neutrinos can be projected onto
the two axes formed by the hadronic tau-decay jets in the transverse
plane, as illustrated in Fig.~\ref{fig:rec}. Knowing the azimuthal
angles of the two $\tau$s, one can determine the two components of
each of the tau neutrino momenta in the transverse plane, and their
momenta in the $z$ direction follow from requiring that the neutrinos
travel in the same direction as the jet. The same recalibration
procedure can be used for $\tau$ pair decays into one or two leptons,
with leptons taking the roles of the hadronic jets, and the sum of two
$\tau$ decay neutrino momenta taking the role of the single $\tau$
decay neutrino of the hadronic case.

In our tau recalibration we assume that there are no other large
contributions to the missing momentum. Possible sources is the
mis-measurement of jet energies or hard neutrinos from the decay of
heavy particles. These will change the direction and magnitude of the
missing momentum vector in the transverse plane. This can partially be
checked for by testing that the direction of the missing momentum lies
inside the opening angle of the two taus.

%%%%%%%%%%%%%%%%%%%%%%%%%%%%%%%%%%%%%%%%%%%%%%%%%%%%%%%%%%%%%%%%%%%%%%%%
\FIGURE[htb]{
\epsfxsize 7.5cm
\epsfbox{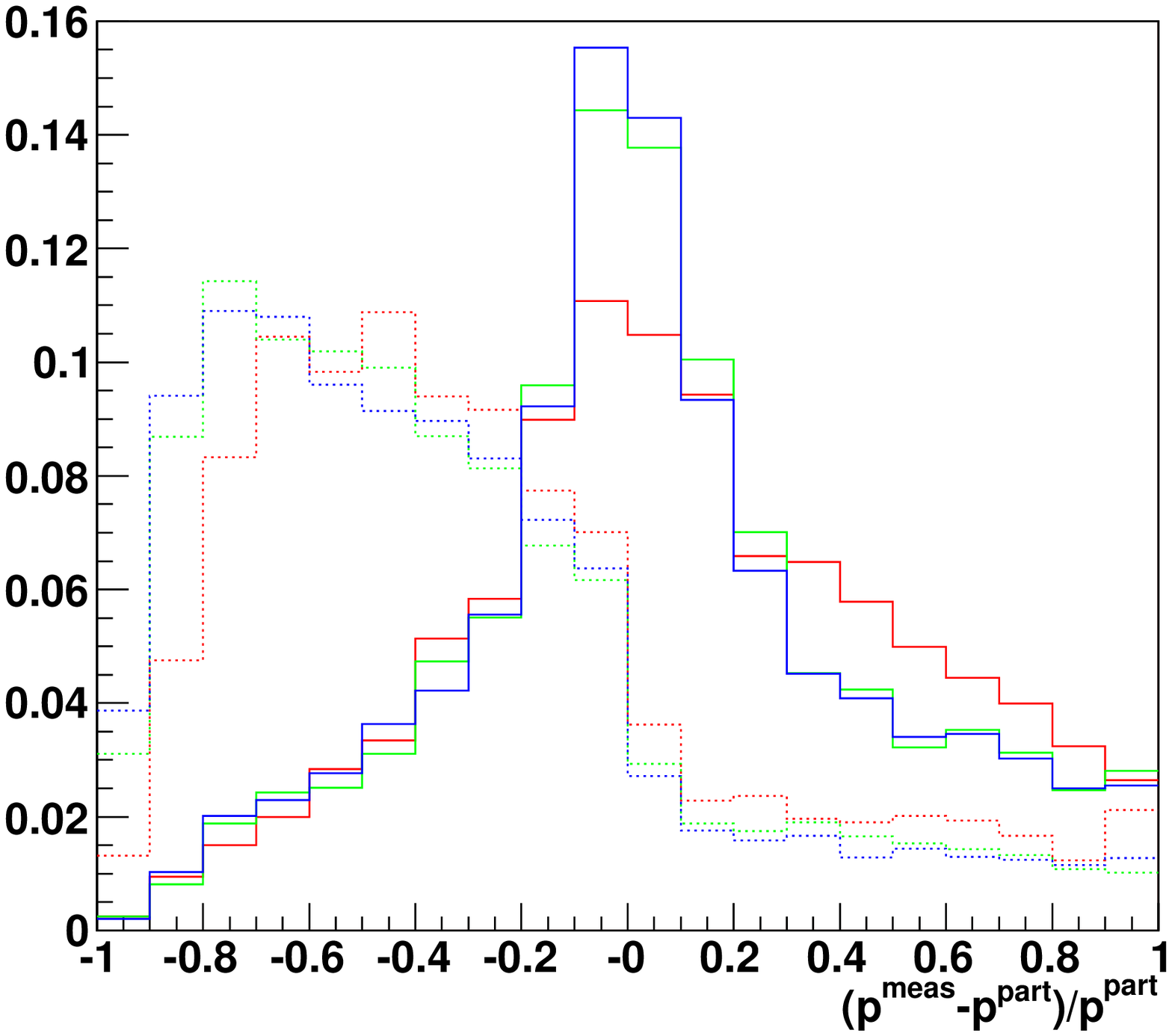}
\caption{The relative error between the parton-level and measured
momenta of $\tau$s at the $\epsilon$ (red), $\zeta$ (green) and
$\eta$ (blue) benchmark points. We show the momenta both before
(dotted line) and after (solid line) recalibration.}
\label{fig:prestau}
}
%%%%%%%%%%%%%%%%%%%%%%%%%%%%%%%%%%%%%%%%%%%%%%%%%%%%%%%%%%%%%%%%%%%%%%

The result of this momentum recalibration is shown in
Fig.~\ref{fig:prestau}, where we plot the relative error on the tau
momentum before (dotted line) and after (solid line) recalibration,
for the three benchmark points. We have used a $50$\% tau-jet tagging
efficiency, and require that for each event the sum of the number of
tagged tau jets and leptons (no staus) with $p_T>10$~GeV is exactly
two, to have an unambiguous pair of tau candidates for the
recalibration. No trigger or background cuts have been used.

We observe a dramatic improvement in the momentum resolution compared
to uncalibrated taus for all three benchmark points, although with a
significant tail of taus with momenta that are too high. The smearing
of the tau momentum is partially due to other contributions to the
momentum imbalance as already mentioned above, leading to a
mis-calibration, but also due to the mis-tagging of tau jets and/or
mis-identification of leptons as coming from tau decays. These
mis-identifications can also be seen in the pre-recalibration
distribution, as a tail to momenta higher than the parton momentum. We
see a marked difference between the $\epsilon$ benchmark and the other
two, the $\epsilon$ benchmark having a worse resolution. This could be
expected from the relatively lower amount of activity in $\zeta$ and
$\eta$ events, where the Drell-Yan and associated production is more
pervasive, resulting in fewer mis-identifications.

%%%%%%%%%%%%%%%%%%%%%%%%%%%%%%%%%%%%%%%%%%%%%%%%%%%%%%%%%%%%%%%%%%%%%%
\subsection{Reconstructing the Decay $\tilde q \to q \tilde\chi_1^0
\to q \tau\tilde\tau_1$}
\label{sect:qRchain}
%%%%%%%%%%%%%%%%%%%%%%%%%%%%%%%%%%%%%%%%%%%%%%%%%%%%%%%%%%%%%%%%%%%%%%

The great majority of right-handed squark, $\tilde q_R$, decays (92
\%) and a few of the left-handed squark, $\tilde q_L$, decays (4 \%)
at the $\epsilon$ benchmark point lead directly to the final state $q
\tilde\chi_1^0$.  To reconstruct the masses of the lightest neutralino
$\tilde\chi_1^0$ and the first two, almost degenerate, generations of
squarks, we seek to isolate events with two squarks, where both decay
according to this clean decay chain~\footnote{We use `clean' here in
the sense that the events contain no hard leptons other than the two
$\tau$s from the $\tilde\chi_1^0$ decay, and only two hard jets,
modulo initial-state radiation (ISR) and final-state radiation (FSR). 
They should therefore provide good measurements
of the transverse missing momenta from the neutrinos in $\tau$
decays.}
\begin{equation}
\tilde q\to q\tilde\chi_1^0\to q \tau\tilde\tau_1.
\label{eq:qR}
\end{equation}
We first select events that survive our trigger requirements and the
standard GDM cuts of Section~\ref{sect:background}. These remove the
Standard Model background, and leave us with an identified pair of
staus. The energies of these staus are then re-calibrated to reflect
the mass measured in Section~\ref{sect:mstau}. Using the parametrised
tau-tagging described in~\cite{heldmann}, with an assumed tau-tagging
efficiency of $50$\%, we select events where the sum of the number of
tau jets with $p_T>20$~GeV and leptons with $p_T>10$~GeV is exactly
two. These are then our basis for reconstructing two taus with the
technique discussed in Section~\ref{sect:tau}. The staus and taus are
matched according to charge, keeping only those events where there is
an unambiguous assignment. This implies that all events with same-sign
staus and all events with two tau jets are rejected. We do not
consider any measurement of tau charge from the hadronic jet, leaving
this as a possible improvement. However, one should note that, while
this will certainly increase the available statistics, the lower
tagging efficiency for hadronic tau decays compared to the efficiency
of identifying leptons means that one would have roughly the same
number of leptonic and hadronic tau candidates. Moreover, the momenta
of the hadronic jets would be known with worse accuracy than those of
the leptons.

Calculating the invariant masses of the $\tilde\tau$-$\tau$ combinations
surviving these cuts we arrive at the distribution shown in the left
plot of Fig.~\ref{fig:qRchain} for the $\epsilon$ benchmark. In blue
we show the distribution due to events with two ${\tilde\chi_1^0} \to
\tilde\tau_1 + \tau$ decays, while the red distribution is the
supersymmetric background. The total distribution shows a clear peak
corresponding to the $\tilde\chi_1^0$ mass. We fit the distribution by
a third-degree polynomial assumption for the background and a
Breit-Wigner distribution for the peak, giving a $\tilde\chi_1^0$ mass
and statistical error of
\begin{equation}
m_{\tilde\chi_1^0} \; = \; 176.6 \pm 0.9~{\rm GeV,}
\label{eq:mchi}
\end{equation}
which is to be compared with the nominal value of $m_{\tilde\chi_1^0}
= 175.2$~GeV. The wide shape of the signal distribution is due to
smearing from the momentum resolution of the staus, and contributes to
the overestimate of the $\tilde\chi_1^0$ mass. For the $\zeta$ and
$\eta$ points the results are very similar, but with significantly
lower statistics, even at $300~\text{fb}^{-1}$. The masses and errors
found are listed in Table~\ref{table:mm}.

%%%%%%%%%%%%%%%%%%%%%%%%%%%%%%%%%%%%%%%%%%%%%%%%%%%%%%%%%%%%%%%%%%%%%%%%
\FIGURE[htb]{
\epsfxsize 15cm
\epsfbox{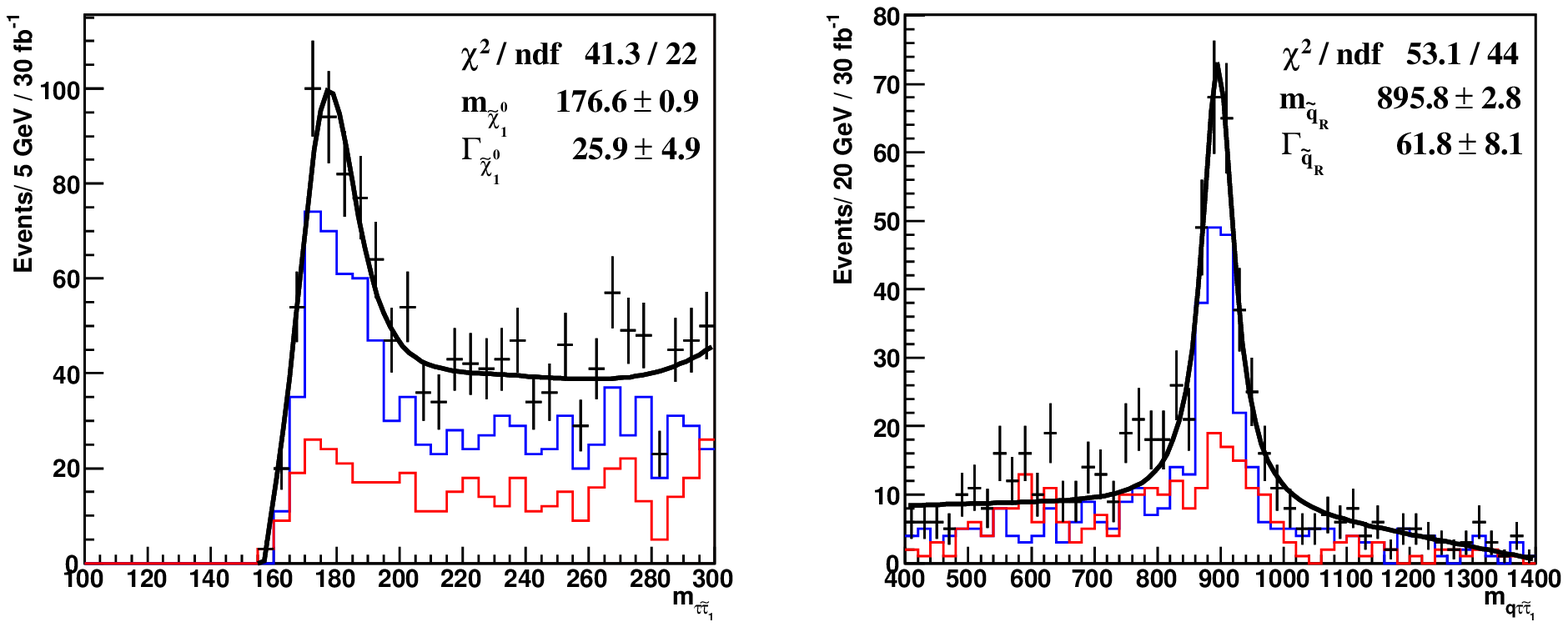}
\caption{The $\tilde\tau$-$\tau$ (left) and jet-$\tilde\tau$-$\tau$ (right)
invariant mass distributions for the $\epsilon$ benchmark point. For
the left plot the blue distribution is for events containing two
decays $\tilde\chi_1^0 \to \tau\tilde\tau_1$, for the right plot
events containing two $\tilde q \to q\tilde\chi_1^0 \to
q\tau\tilde\tau_1$ decay chains. The red distributions include all
other events surviving the cuts described in the text.}
\label{fig:qRchain}
}
%%%%%%%%%%%%%%%%%%%%%%%%%%%%%%%%%%%%%%%%%%%%%%%%%%%%%%%%%%%%%%%%%%%%%%

To find the (right-handed) squark mass, we want to isolate events that
contain two of the decays of Eq.~(\ref{eq:qR}). This is done by adding
to the cuts used above the following:
\begin{itemize}
\item
We require two jets with $p_T>300,150$~GeV,
\item
no other jets with $p_T>150$~GeV,
\item
missing transverse energy $E_T<50$~GeV, and
\item
to increase statistics we relax the unambiguous assignment requirement
for the $\tau$-$\tilde\tau$ pairs, keeping those not directly excluded
by charge considerations and which have an invariant mass within two
times the peak width of the $\tilde\chi^0_1$,
$\Gamma_{\tilde\chi^0_1}=25.9$~GeV.
\end{itemize}
Here the cut on missing energy attempts to remove events with
additional hard neutrinos, not coming from the decay of the final taus
in the decay chain.

The two hardest jets in the event are then combined with the
$\tilde\tau$-$\tau$ pairs and the invariant masses of the triples are
found. The invariant mass distribution is shown on the right in
Fig.~\ref{fig:qRchain}. The surviving events
are very predominantly decays of the $\tilde q_R$, and
the total distribution again shows a clear
peak, now corresponding to the $\tilde q_R$ mass~\footnote{As we show
below, the $\tilde q_L$ mass can be measured using a different decay
chain.}. Fitting the distribution with a second-degree polynomial for
the background plus a Breit-Wigner, at the $\epsilon$ benchmark point
we find
\begin{equation}
m_{\tilde q_R} \; = \; 895.8 \pm 2.8~{\rm GeV}.
\label{eq:mqR}
\end{equation}
This can be compared to the nominal value of $m_{\tilde q_R} =
900.8$~GeV at the $\epsilon$ benchmark point. The background tail to
wards lower invariant masses tends to result in a small underestimate
of the mass. We note that the accuracy of the squark mass measurement
is dependent on the jet recalibration routine used. The statistical
error after three years at low luminosity is lower than the
systematical error expected from the uncertainty in the absolute jet
energy scale, which is estimated to be of the order of $1$\% (see
Chapter 12 of~\cite{:1999fq}). For the $\zeta$ benchmark point we
again have a clear excess of events over background, and we can
reconstruct a mass peak, but only containing a few dozen events. The
statistical error for $\zeta$ is large, and varies somewhat with
binning and fit range. For $\eta$ only two signal events survive all
cuts, compared to a background of ten, thus we are not able to
reconstruct the $\tilde q_R$ in this case. We list the numerical results in
Table~\ref{table:mm}.

By using jet-$\tilde\tau$-$\tau$ combinations where the jet is tagged as
coming from a $b$ quark, we can estimate the $\tilde b$ mass. Because
of the low statistics obtainable with $30~\text{fb}^{-1}$ of
luminosity we can provide only a rough estimate of a combined $\tilde
b_1$ and $\tilde b_2$ mass. Assuming a $53$\% b-tagging efficiency,
with light jet rejection as discussed in Section~\ref{sect:MC}, we
have
\begin{equation}
m_{\tilde b} \; = \; 888.9\pm 57.7~{\rm GeV}
\label{bmass}
\end{equation}
We note that, whilst the production cross sections of the $\tilde b_1$
and $\tilde b_2$ are of the same magnitude, the branching ratio of
$\tilde b_2 \to b\tilde\chi_1^0$ is much larger than that of $\tilde
b_1$, as $\tilde b_2$ is mostly $\tilde b_R$. Consequently, events with
$\tilde b_2$ dominate this mass measurement. We recall that $m_{\tilde
b_2} = 883$~GeV at the benchmark point $\epsilon$. For the other two
benchmark points we were unable to measure the $b$-squark masses due
to the low number of events.

Almost a half of the sparticle pair-production events at the LHC at
benchmark point $\epsilon$ would include gluinos, and the great
majority of their decays would be into squark-quark pairs. One may
therefore hope to reconstruct a $\tilde g$ mass peak by plotting the
invariant masses of jet-jet-$\tilde\tau$-$\tau$ combinations. However,
if we take the events lying within twice the width of the squark mass
peak found here, and add a second jet to each jet-$\tilde\tau$-$\tau$
combination we find that while there is a clear peak at the gluino
mass in events which contain the decay of Eq.~(\ref{eq:qR}), initiated
by a gluino, the supersymmetric background has a similar shape at
roughly the same position and is of the same magnitude, thus
preventing any gluino mass determination. This can be understood in
terms of the softer nature of the jets from the gluino decay as
compared to the squark decay; the background jets have a very similar
kinematic distribution. One possible idea, which is beyond the scope
of this paper to explore further, is the decay $\tilde g
\to t \tilde{t}_1 \to tt \tilde\chi_1^0$. This could be used to find the
gluino mass if the top decays involved could be fully
reconstructed. The decay has the advantage of a large branching ratio,
but reconstructing the $W$ involved from hadronic decays will present
a formidable challenge. For the other two benchmark points, although
the jets from gluino decays are somewhat harder, the background
problem still remains.

It might be possible to improve on the above results by refining the
event selection, e.g., by using only leptonic $\tau$ decays for high
statistics scenarios, and/or by making a more sophisticated analysis
of the final states, for example by modelling the expected
supersymmetric background in a more sophisticated way than our
polynomial assumptions. However, the results in Fig.~\ref{fig:qRchain}
and Eqs.~(\ref{eq:mchi}) and (\ref{eq:mqR}) already improve
significantly on those shown in~\cite{DeRoeck:2005bw}, where no
attempt was made to correct for the missing neutrino momenta in $\tau$
decays, and the full kinematics of the decay was not exploited. We
also observe that for the $\epsilon$ benchmark point our statistical
errors are in general below the expected levels of errors from
systematical uncertainties in lepton and jet energy scales at the LHC.
%The results of the simulations for all three benchmark points are
%summarised in Table~\ref{table:mm}.

%%%%%%%%%%%%%%%%%%%%%%%%%%%%%%%%%%%%%%%%%%%%%%%%%%%%%%%%%%%%%%%%%%%%%%
\subsection{Reconstructing the Decay $\tilde q \to q \tilde\chi_2^0
\to q \ell \tilde{\ell} \to q \ell\ell \tilde\chi_1^0 \to q\ell\ell
\tau\tilde\tau_1$}
\label{sect:qLchain}
%%%%%%%%%%%%%%%%%%%%%%%%%%%%%%%%%%%%%%%%%%%%%%%%%%%%%%%%%%%%%%%%%%%%%%
Although the long decay chain
\begin{equation}
\tilde q \to q \tilde\chi_2^0 \to q \ell \tilde{\ell}
\to q \ell\ell \tilde\chi_1^0 \to q\ell\ell\tau\tilde\tau_1,
\label{eq:qL}
\end{equation}
predominantly of the left-handed squark via a left-handed slepton, has
a much lower branching ratio than the decay of the right-handed squark
discussed in the previous Section, it has a distinctive signature of
between two and four leptons, depending on the $\tau$ decay modes, 
in addition to the pair of $\tilde\tau$s.

%%%%%%%%%%%%%%%%%%%%%%%%%%%%%%%%%%%%%%%%%%%%%%%%%%%%%%%%%%%%%%%%%%%%%%%%
\FIGURE[htb]{
\epsfxsize 7.5cm
\epsfbox{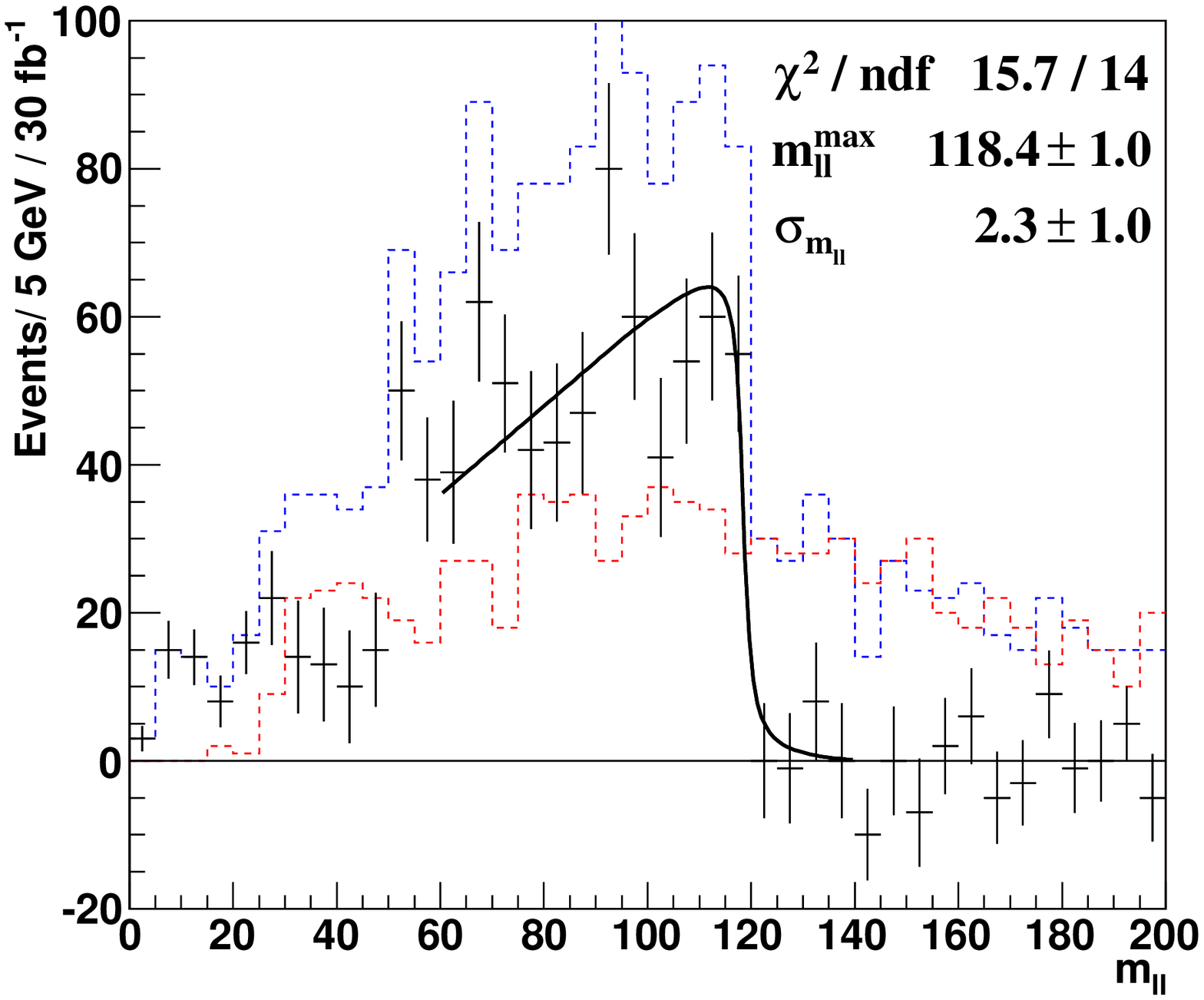}
\caption{The different-flavour subtracted dilepton invariant mass
distribution for the $\epsilon$ benchmark point. We also show the
opposite-sign, same-flavour distribution (blue, dashed line) and the
opposite-sign, different-flavour distribution (red, dashed line).
\label{fig:mll}}
}
%%%%%%%%%%%%%%%%%%%%%%%%%%%%%%%%%%%%%%%%%%%%%%%%%%%%%%%%%%%%%%%%%%%%%%

It is well known that a very useful step in the reconstruction of
analogous decay chains in neutralino dark matter scenarios is the
measurement of the dilepton spectrum resulting from $\tilde\chi_2^0
\to \ell\ell \tilde\chi_1^0$ decays. Accordingly, we start here by
also looking at the dilepton edge for the $\epsilon$ benchmark point,
plotting the invariant mass of pairs of same-flavour, opposite-sign
leptons in the MC data. After the standard GDM cuts from
Section~\ref{sect:MC} to isolate events with two staus, loosening the
cut on tau candidates to allow for more leptons, we pick events with a
minimum of two additional leptons, each having $p_T>30$
GeV.\footnote{We have checked that the generated SM background
discussed in Section~\ref{sect:background} remains zero after these
changes.} Combining all pairs of these leptons with the same flavour
and opposite sign we arrive at the blue distribution in
Fig.~\ref{fig:mll}.  We can remove the background of uncorrelated
leptons by making the standard subtraction of the same flavour
distribution (red) under the lepton universality assumption. The
resulting, well known, triangular-shaped distribution from the decay
$\tilde\chi_2^0 \to \ell \tilde{\ell} \to \ell\ell\tilde\chi_1^0$
appears, and we can perform a fit to the endpoint.  Using a straight
line fit, smeared by a Gaussian distribution to simulate sparticle
finite-width effects and detector smearing at the edge, we measure
\begin{equation}
m_{\ell\ell}^{\max} \; = \; 118.6 \pm 1.0~{\rm GeV},
\label{eq:mll}
\end{equation}
where $m_{\ell\ell}^{\max}$ is the position of the di-lepton edge, given in
terms of the masses of the sparticles in the decay chain by the
following formula:
\begin{equation}
(m_{\ell\ell}^{\max})^2 = \frac{(m_{\tilde\chi_2^0}^2-m_{\tilde{\ell}}^2)(m_{\tilde{\ell}}^2-m_{\tilde\chi_1^0}^2)}{m_{\tilde{\ell}}^2}.
\label{eq:mlldef}
\end{equation}
The nominal value for the $\epsilon$ benchmark is
$m_{\ell\ell}^{\max}=119.7$~GeV. For $\zeta$ and $\eta$ the endpoints
are found at $m_{\ell\ell}^{\max}=332.2\pm 9.0$~GeV and
$m_{\ell\ell}^{\max}=339.1\pm 9.8$~GeV, respectively.

With this information in hand we start the search for the decay in
(\ref{eq:qL}). First we attempt a reconstruction of the slepton mass
in the decay $\tilde{\ell} \to \ell \tilde\chi_1^0 \to \ell
\tau\tilde\tau_1$, using the standard GDM cuts of 
Section~\ref{sect:background} with some modifications:
\begin{itemize}
\item
We allow for additional leptons to the tau candidates, iterating over
the assigment of leptons as tau candidates or as ``additional'',
\item
we require that we have no more than two tau-tagged jets to avoid
neutrinos from other tau decays, and
\item
the tau candidates are paired with staus, and we keep only events
where there is at least one pair with consistent charge and invariant
mass within two times the width of the $\tilde\chi_1^0$ mass peak, as
measured in Section~\ref{sect:qRchain}.
\end{itemize}
Following these cuts we add the additional lepton(s) to the
reconstructed $\tilde\chi_1^0$, and show the resulting invariant mass
distribution of the lepton-$\tau$-$\tilde\tau$ triples in the left
plot of Fig.~\ref{fig:qLchain}. After fitting the distribution with an
assumed linear background and Breit-Wigner resonance we find a slepton
mass of
\begin{equation}
m_{\tilde \ell} \; = \; 307.9 \pm 2.0~{\rm GeV,}
\label{eq:mslepton}
\end{equation}
which can be compared with the nominal value of $m_{\tilde
\ell_L}=305.9$~GeV.

%%%%%%%%%%%%%%%%%%%%%%%%%%%%%%%%%%%%%%%%%%%%%%%%%%%%%%%%%%%%%%%%%%%%%%%%
\FIGURE[htb]{
\epsfxsize 15cm
\epsfbox{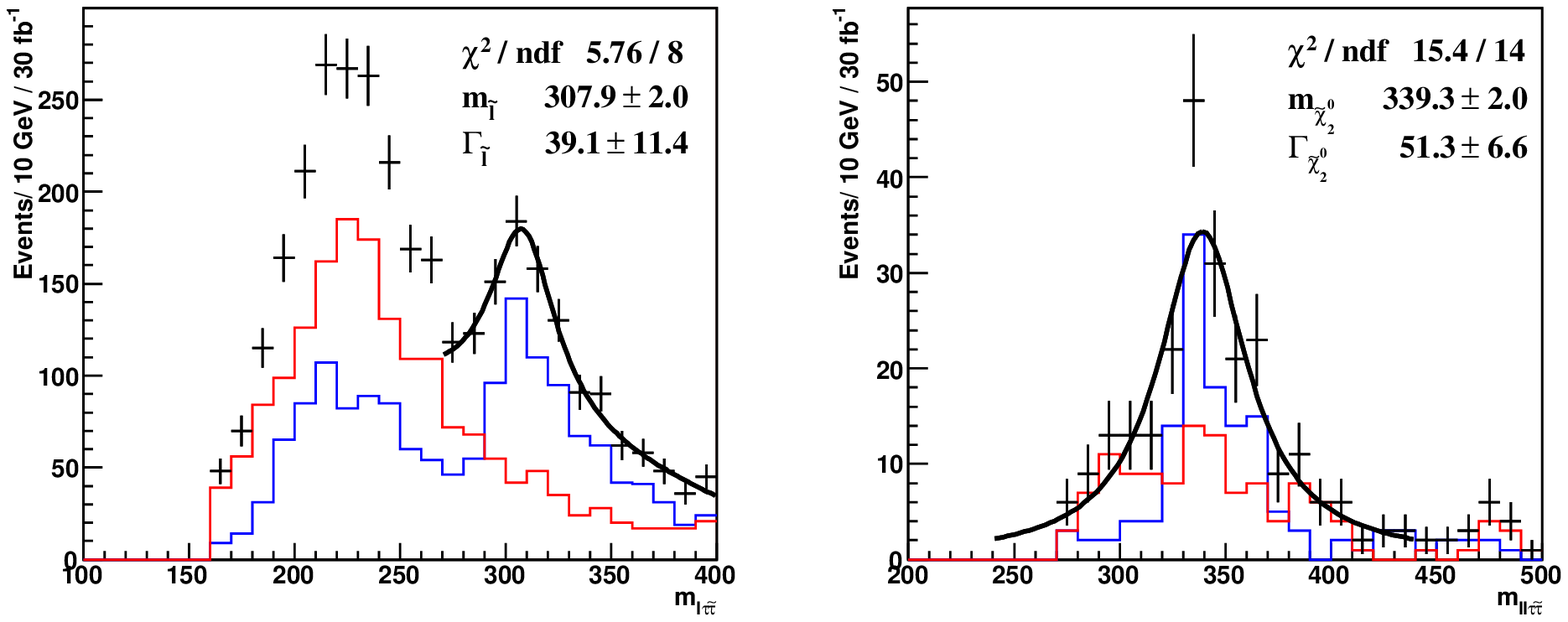}
\caption{The lepton-$\tilde\tau$-$\tau$ (left) and
lepton-lepton-$\tilde\tau$-$\tau$ (right) invariant-mass distributions
for the $\epsilon$ benchmark point. The blue distributions are for
signal events, events containing a decay
$\tilde{\ell}\to\ell\tilde\chi_1^0 \to \ell\tau\tilde\tau_1$ (left)
and $\tilde\chi_2^0\to \ell \tilde{\ell} \to \ell\ell \tilde\chi_1^0
\to \ell\ell\tau\tilde\tau_1$ (right), while the red distributions
include all other events that pass the cuts given in the text.
\label{fig:qLchain}}
}
%%%%%%%%%%%%%%%%%%%%%%%%%%%%%%%%%%%%%%%%%%%%%%%%%%%%%%%%%%%%%%%%%%%%%%

The peak in the blue distribution to the left of the resonance that
can be seen in Fig.~\ref{fig:qLchain} is due to the combination of
$\tilde\chi_1^0$ candidates with the softer lepton coming from
$\tilde\chi_2^0$ decays, while the peak in the red distribution is due
to generic soft leptons added to the $\tilde\chi_1^0$ resonance. From
this we suspect that a scenario with a small $m_{\tilde
l}-m_{\tilde\chi_1^0}$ mass difference will encounter the same
problems as the gluino mass measurement, while the $\tilde\chi_2^0$
mass will still be easily accessible. In such scenarios there will be
an advantage in first measuring the $\tilde\chi_2^0$ mass.

For the other two benchmark points we get similar distributions, but
again of course with less events and fitted masses with larger
statistical errors. For $\zeta$ we get $m_{\tilde \ell}=674.6\pm
3.4$~GeV, and for $\eta$, $m_{\tilde\ell}=673.5\pm 3.5$~GeV. We note
that the value for $\zeta$ is quite some distance from the nominal
value of $m_{\tilde\ell}=683.8$~GeV. The distribution of the resonance
peak is very broad and extends towards lower invariant masses from a
maximum around the true maximum.

Starting from the slepton resonance we make the following additional
cuts to find the mass of the $\tilde\chi_2^0$ for the $\epsilon$
benchmark point:
\begin{itemize}
\item
We take events with lepton-$\tilde\tau$-$\tau$ combinations within
twice the width of the slepton mass peak,
\item
we require events to have at least two leptons in addition to the tau
candidates,
\item
the additional leptons are required to have opposite sign and same
flavour, and to have an invariant mass $m_{\ell\ell}<120$~GeV, on the basis
of the measured dilepton edge.
\end{itemize}
Taking all allowed lepton pairs in an event and plotting the
invariant mass of the lepton-lepton-$\tilde\tau$-$\tau$ combinations,
we arrive at the distribution in the right-hand plot of
Fig.~\ref{fig:qLchain}. Fitting the clear mass peak with a
Breit-Wigner function we find a $\tilde\chi^0_2$ mass of
\begin{equation}
m_{\tilde\chi^0_2} \; = \; 339.3 \pm 2.0~{\rm GeV},
\label{eq:mchi02}
\end{equation}
to be compared with the nominal value $m_{\tilde\chi^0_2}=339.4$~GeV.

For the $\zeta$ and $\eta$ points, the statistics are very low after all
cuts. We find no real mass peak for $\zeta$, with only a handful of
events scattered over a large mass interval, and a very broad
resonance for $\eta$, from which one can estimate
$m_{\tilde\chi^0_2}=800 \pm 50$~GeV. However, from the formula for the
dilepton edge in Eq.~(\ref{eq:mlldef}) and the $\tilde\chi^0_1$ mass
found in the previous Section we can calculate the $\tilde\chi^0_2$
mass and statistical error for $\zeta$ and $\eta$. The results are
listed in Table~\ref{table:mm}.

The information from the dilepton edge could also be used to restrict further
the masses at the $\epsilon$ benchmark by a combined $\chi^2$
fit, but the effect is small, although the statistical error on
the $\tilde\chi^0_2$ mass can be halved. To increase the accuracy of
the mass determination, in particular for the benchmark points with
low cross sections, it could also be interesting to pursue the other
invariant mass distributions that can be constructed from the leptons
in this decay chain, e.g., the lepton-stau invariant masses and the
lepton-lepton-stau invariant mass. A generalisation of the formulae of
\cite{Miller:2005zp}, which describe the shapes of invariant mass
distributions in cascade decays, to include massive stable
end-products like the stau is also possible. This is, however, outside
the scope of this initial study.

What remains from the decay chain in (\ref{eq:qL}) is to determine the
left-handed squark mass. Taking the events in the $\tilde\chi_2^0$
mass peak, we add the two hardest jets of the event to the
reconstructed $\tilde\chi_2^0$. Because there is a significant
fraction of background events, mostly events where the
$\tilde\chi_2^0$ does not come from the decay of a left-handed squark,
we additionally require a missing transverse energy
$E_T<50$~GeV. While statistics are very low as a result of this
selection, we find a mass peak and can perform a fit with a
Breit-Wigner distribution. The result is plotted in Fig~\ref{fig:qL},
and we find a left-handed squark mass of
\begin{equation}
m_{\tilde q_L} \; = \; 923.4 \pm 3.7~{\rm GeV},
\label{eq:mqL}
\end{equation}

%%%%%%%%%%%%%%%%%%%%%%%%%%%%%%%%%%%%%%%%%%%%%%%%%%%%%%%%%%%%%%%%%%%%%%%%
\FIGURE[htb]{
\includegraphics[width=7.4cm]{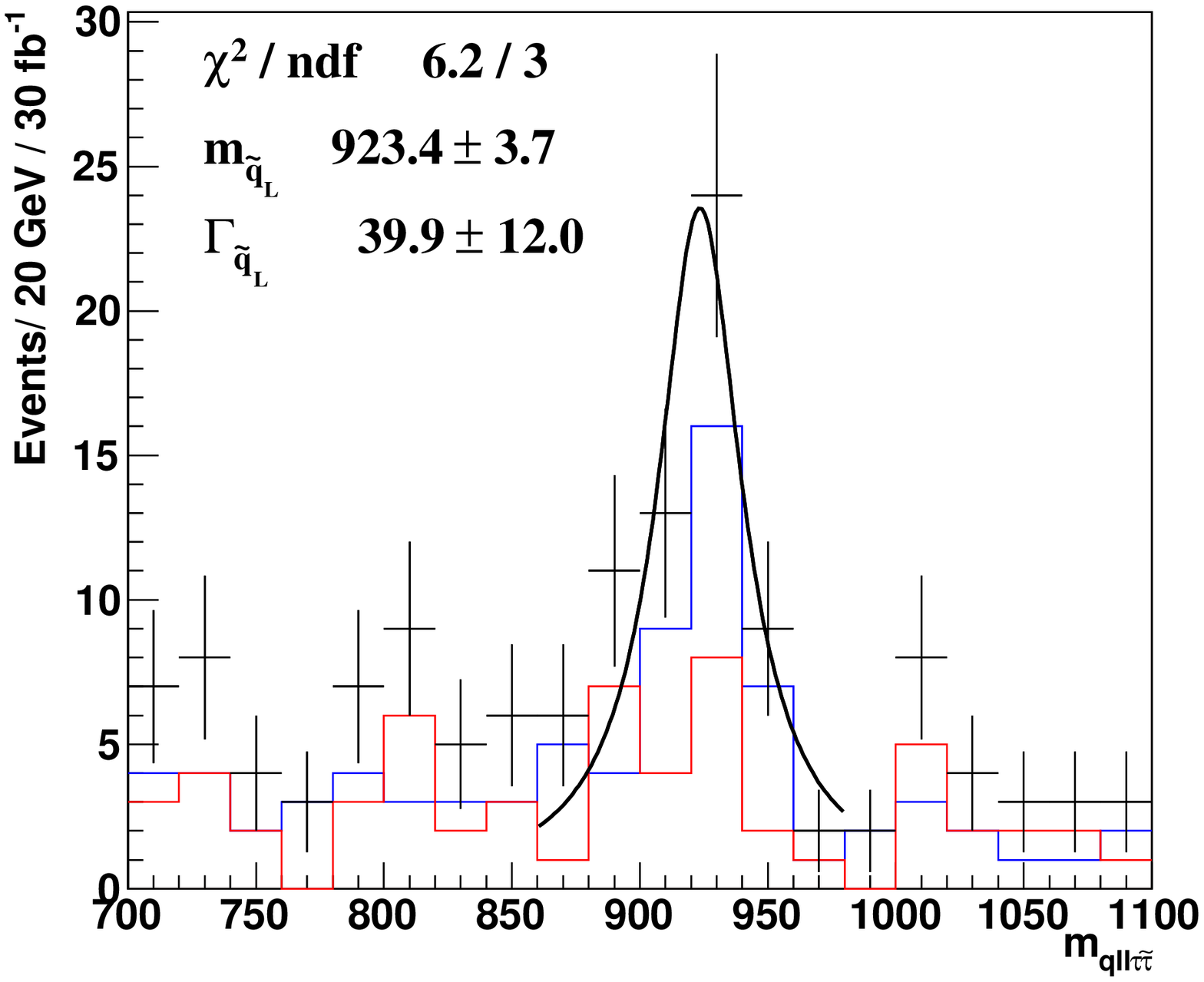}
\includegraphics[width=7.4cm]{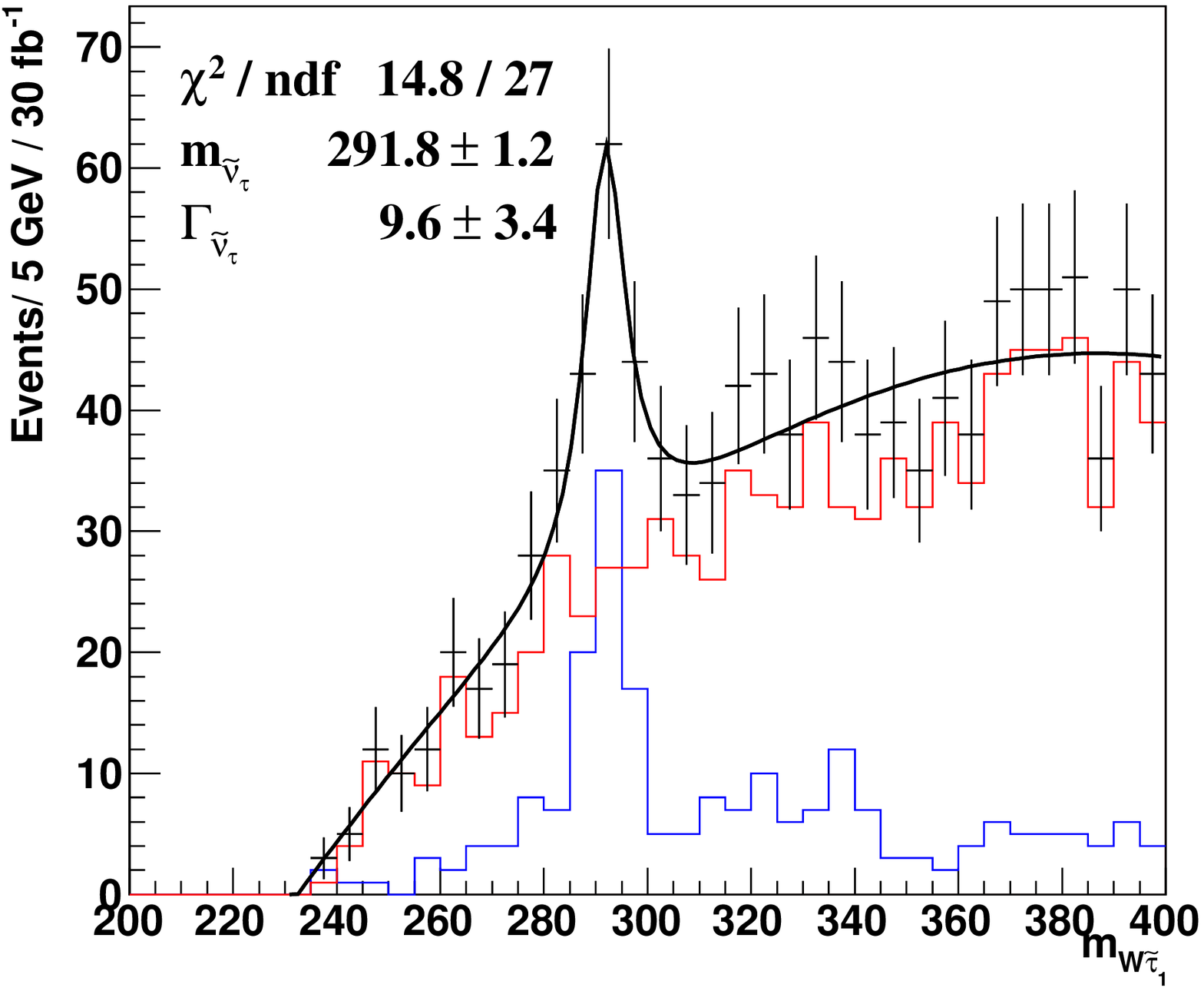}
\caption{The jet-lepton-lepton-$\tilde\tau$-$\tau$ (left) and the
$W$-$\tilde\tau$ (right) invariant-mass distributions for the
$\epsilon$ benchmark point. The blue distribution is for signal
events, events containing the decays of Eq.~\ref{eq:qL} and
$\tilde\nu_\tau\to W\tilde\tau_1$, respectively, while the red
distributions include all other events that pass the cuts given in the
text.
\label{fig:qL}}
}
%%%%%%%%%%%%%%%%%%%%%%%%%%%%%%%%%%%%%%%%%%%%%%%%%%%%%%%%%%%%%%%%%%%%%%

\noindent
Compared to the common mass of the first two generations of both
families, $m_{\tilde d_L, \tilde s_L}=938.2$~GeV and $m_{\tilde u_L,
\tilde c_L}=934.5$ GeV, this fit result is slightly low, indicating
that there may be some systematical effect, and with a relative error
of the order of 1\%. However, comparing with the squark mass found in
the previous Section, this analysis clearly demonstrate that there are
two different masses and the decay chains allow us to assign the left-
and right-handedness as we have done. Since we could not reconstruct a
$\tilde\chi_2^0$ mass peak, the $\tilde q_L$ mass is of course also
inaccessible with this technique at the $\zeta$ and $\eta$ benchmarks.

%%%%%%%%%%%%%%%%%%%%%%%%%%%%%%%%%%%%%%%%%%%%%%%%%%%%%%%%%%%%%%%%%%%%%%
\subsection{Reconstructing $\tilde\nu_\tau$ and $\tilde\chi^\pm_1$ Decays}
\label{sect:associated}
%%%%%%%%%%%%%%%%%%%%%%%%%%%%%%%%%%%%%%%%%%%%%%%%%%%%%%%%%%%%%%%%%%%%%%
The high fraction of the total cross section that comes from the
associated production of $\tilde\chi^\pm_1\tilde\chi^0_2$ pairs at the
$\zeta$ and $\eta$ benchmark points, together with the large
production of neutralinos and charginos in the decays of squarks at
all three points, makes the dominant decays of $\tilde\chi^\pm_1$ 
potentially as interesting as the $\tilde\chi^0_2$ decay via a left-handed
slepton that was discussed in the previous Section.
Accordingly, we look now at the corresponding decays of the chargino.

In the case of the $\epsilon$ benchmark point, the decay chain
$\tilde\chi^\pm_1\to\tau\tilde\nu_\tau\to\tau W\tilde\tau_1$ has a
total branching ratio of $9.5$\%, and the decay
$\tilde\chi^0_2\to\nu_\tau\tilde\nu_\tau\to\nu_\tau W\tilde\tau_1$, a
branching ratio of $9.0$\%. This makes the search for a
$\tilde\nu_\tau$ resonance feasible, if we can reconstruct hadronic
decays of the $W$ in the supersymmetric events.

The reconstruction of $W$s produced in the decays of heavy particles
at the LHC is difficult because of the large jet background in the
high multiplicity environment. Additionally, since $W$s will be so
copiously produced in SM events at the LHC, it will be difficult to
find effective background cuts. When a $W$ is produced with little
boost in the laboratory frame, the jets from the decay are relatively
soft and easily drown among the other jets of the event.  For boosted
$W$s the jets are harder, but have a much smaller opening angle in the
lab frame, and are thus easily mistaken for a single jet by a cone jet
algorithm. The quest for possible improvements in the search for
hadronic decays of heavy bosons in supersymmetric events by the use of
different jet algorithms is an interesting topic, but lies beyond the
scope of this paper. As we shall see, because of the excellent
rejection of the Standard Model backgrounds afforded by the properties
of the stau, the identification of $W$ candidates is easier in the GDM
scenarios.

To isolate events with the decay $\tilde\nu_\tau\to W\tilde\tau_1$, we
apply the standard GDM cuts of Section~\ref{sect:MC}, with the
exception of the requirement of two tau candidates. Instead we tighten
the cut on stau velocity to $\beta\gamma<2.7$, which we again find
removes all SM background. This allows us a very simple identification
of events with $W$s. We require one and only one combination of two
jets in the event which has:
\begin{itemize}
\item
Invariant mass within two times the width, $\Gamma_W=2.12$~GeV, of the
$W$ mass, $m_W=80.42$~GeV, and
\item
an opening angle $\theta<\frac{\pi}{4}$ in the laboratory frame.
\end{itemize}
The cut on angle is effective in rejecting the large background of
events without a $W$, where by chance two relatively soft jets have the
correct invariant mass, while, for reasons discussed above, it is a
typical property of $W$s that come from the decays of heavier
sparticles. After these cuts and after combining the reconstructed $W$
candidates with the closest stau we arrive at the invariant mass
distribution shown in Fig~\ref{fig:qL}. The signal distribution (blue)
shows a clear peak above the background (red), which consists mostly
of events without a $W$. By fitting the total distribution with the
sum of a third-degree polynomial for the background and a Breit-Wigner
distribution for the peak, we arrive at a value of the tau-sneutrino
mass of
\begin{equation}
m_{\tilde\nu_\tau} \; = \; 291.8 \pm 1.2~{\rm GeV}
\label{eq:msnutau}
\end{equation}
to be compared with the nominal value of
$m_{\tilde\nu_\tau}=288.5$~GeV. The overestimated value for $\epsilon$
is due to the long tails of the signal distribution to the right of
the peak, which in turn comes from combinatorial effects, such as
pairing the $W$ with the wrong stau candidate. 

From events in the tau-sneutrino resonance peak we should in principle
be able to reconstruct the chargino mass by adding tau jets to the
reconstructed sneutrinos. However, the large background and the soft
nature of these taus means that the signal drowns in a background of
the same shape and position, much as in the case of the
gluinos. For the $\zeta$ and $\eta$ benchmark points the situation is
similar, we can reconstruct a $\tilde\nu_\tau$ mass peak, but no
chargino. The sneutrino masses found for these points are listed in
Table~\ref{table:mm}.

%%%%%%%%%%%%%%%%%%%%%%%%%%%%%%%%%%%%%%%%%%%%%%%%%%%%%%%%%%%%%%%%%%%%%%
\section{Conclusions}
\label{sect:conclusion}
%%%%%%%%%%%%%%%%%%%%%%%%%%%%%%%%%%%%%%%%%%%%%%%%%%%%%%%%%%%%%%%%%%%%%%
We have demonstrated in this paper that the LHC detectors (as
exemplified here by the ATLAS detector), would have the abilities to
make many valuable measurements in benchmark supersymmetric scenarios
with gravitino dark matter and a $\tilde\tau$ NLSP. The triggers
planned would record GDM events with high redundancy and efficiency,
and the robust designs of the detectors would enable them to
characterise very well $\tilde\tau$s produced in LHC collisions. Once
recorded, the $\tilde\tau$ tracks in GDM events could be identified
with high efficiency, and useful momentum, ToF and $dE/dx$
measurements would be made. We find that it should be possible to
measure the stau mass with statistical and systematic errors at the
sub-permille level, assuming only the reconstruction of staus with
velocity down to $\beta\gamma\approx 0.6$. The majority of GDM events
with a pair of $\tilde\tau$ NLSPs would also contain a pair of $\tau$
leptons, which should be detectable with good efficiency at the
LHC. Combining these with the $\tilde\tau$ measurements would enable
first the $\tilde\chi^0_1$ and subsequently heavier sparticles to be
reconstructed and their masses measured with accuracies that often
approach the expected systematic energy scale uncertainty of jets and
leptons. We demonstrate these abilities with the aid of $\tau$
momentum recalibration using transverse momentum balance, and making
appropriate kinematic cuts on the final-state kinematic distributions
in sparticle cascade decays. Our results are consistent with the
preliminary estimates of sparticle observability made
in~\cite{DeRoeck:2005bw}.

The LHC will be the first accelerator to explore a new energy domain,
and there are high hopes that it will reveal new physics beyond the
Standard Model. Supersymmetry is among the most `expected' new physics
surprises, which has been much studied by the LHC collaborations as
well as by theorists. Most of these studies have assumed the
`expected' missing-energy signature of R-conserving supersymmetry with
a neutralino LSP, though other possibilities have also been
explored. It has been realised recently that some of these other
possibilities may be as generic as missing energy, and the scenario
studied here of gravitino dark matter and a stau NLSP is just one of
many possible `unexpected' possibilities. Nevertheless, it is
encouraging that, although the LHC detectors were not designed with
such an exotic scenario explicitly in mind, they seem to be capable
also of dealing with such an `unexpected' physics surprise. We have
shown here that valuable insight into such a gravitino dark matter
scenario could be obtained already with a relatively modest initial
amount of LHC luminosity, and so could inform the design of possible
future experiments at the high-energy frontier.

\acknowledgments
We thank Giacomo Polesello for helpful discussions on a number of
topics. ARR acknowledges support from the European Community through a
Marie Curie Fellowship for Early Stage Researchers Training.

%%%%%%%%%%%%%%%%%%%%%%%%%%%%%%%%%%%%%%%%%%%%%%%%%%%%%%%%%%%%%%%%%%%%%%

\end{document}